\documentclass{aa}             
\usepackage{aabib99}
\usepackage{graphics}
\usepackage{times}

\begin{document}

\thesaurus{
	      09         
	       09.03.1;  
	       09.08.1;  
	       09.11.1;  
	       09.13.2;  
	     }
\title{ A ridge of recent massive star formation between Sgr B2M 
and Sgr B2N}
\subtitle{ }
\author{ P. de Vicente\inst{1}, J. Mart\'\i n-Pintado\inst{1}, 
R. Neri\inst{2} \and P. Colom\inst{3}}
\institute{ Observatorio Astron\'omico Nacional,
	    Apartado 1143,
	    28080 Alcal\'a de Henares,
	    Spain\and
            Institut de Radioastronomie Millimetrique
            Rue de la Piscine, 
            St Martin de Heres
            France\and
            Observatoire de Meudon
            Meudon,
            Paris
            France}
\offprints{P. de Vicente}
%
\date{ Received, 2000/ accepted,  2000}
%
\authorrunning{de Vicente et al.}
\maketitle
\begin{abstract}

We present single dish and interferometric maps of several rotational 
transitions of HC$_3$N vibrationally excited levels towards Sgr B2. 
HC$_3$N is a very suitable molecule to probe hot and dense regions (hot cores) 
affected by high extinction since
its vibrational levels are mainly excited by mid-IR radiation. The single dish 
maps show, for the first time, that the HC$_3$N vibrationally 
excited emission (HC$_3$N*) is not restricted to Sgr B2M and N but extended 
over an area $40''\times20''$ in extent. We 
distinguish four bright clumps (Sgr B2R1 to B2R4) in the ridge connecting 
the main cores Sgr B2M and Sgr B2N, and a low brightness extended region 
to the west of the ridge (Sgr B2W). The physical properties and the kinematics 
of all hot cores have been derived from the HC$_3$N* 
lines. Our high angular resolution images show that the Sgr B2N hot core
breaks in two different hot cores, Sgr B2N1 and N2, with different radial
velocities and separated by $\sim 2"$ in declination.
We find that the excitation of the HC$_3$N*  emission in
all hot cores can be represented by a single temperature
and that the linewidth of the HC$_3$N* rotational lines arising from different
vibrational levels systematically decreases as the energy of the vibrational
level increases. The systematic trend in the linewidth is likely related to 
the increase of the velocity as the distance to the exciting source increases.
We have developed a simple model to study the excitation
of the HC$_3$N vibrational levels by IR radiation.  We find that the single
excitation temperature can be explained by high luminosities 
of embedded stars 
($\sim 10^7$ L$_\odot$ ) and small source sizes ($\sim 2-3''$).
The estimated clump masses are 500 M$_\odot$ for Sgr B2M, 
800 M$_\odot$ for Sgr B2N
and 10-30 M$_\odot$ for Sgr B2R1 to B2R4. Luminosities are $1-2 \, 10^6$ 
L$_\odot$ for Sgr B2R1-B2R4 and Sgr B2M and $10^7$ L$_\odot$ for Sgr B2N. 
We estimate HC$_3$N abundances of  $5\, 10^{-9}$ for Sgr B2M and 
Sgr B2N2 and 
$10^{-7}$ for the rest of the hot cores. The different HC$_3$N abundances 
in the hot cores reflect different stages of evolution due to time
dependent chemistry and/or photo-dissociation by UV radiation from nearby HII
regions.
According to the mass and the luminosity of the different hot cores, we 
propose that Sgr B2M and B2N contain a cluster of 20-30 hot cores, each
like that in Orion A, a number similar to the UC HII regions already 
detected in the region. The Sgr B2R1-B2R4 hot cores represent isolated 
formation of massive stars.


\end{abstract}
\section{Introduction}

Recent massive star formation in the Galaxy is commonly recognized by 
signposts, like UC HII regions, H$_2$O masers, near and mid 
infrared emission and 
hot and dense condensations containing molecules and dust.
These condensations are known as hot cores and 
represent a relatively early phase of massive star formation. 

The Sagittarius B2 molecular cloud, located near the Galactic center, at
a distance of 7.1 kpc \cite{reid88} is 
one of the most active regions of massive star formation in the Galaxy.
It contains two cores, Sgr B2M and Sgr B2N, which show all the typical 
signposts of very luminous embedded young stellar objects.
Sgr B2M and Sgr B2N, are
very strong IR emitters \cite{thronson86,goldsmith87,goldsmith90}, 
they contain several compact and ultra compact HII
regions \cite{gaume90,gaume95,depree98}, 
hot cores \cite{vogel87,mehringer97,martin-pintado90,devicente97} and 
H$_2$O masers \cite{reid88,kobayashi89}. 

 The Sgr B2N and Sgr B2M hot cores have already been studied for example,
 in NH$_3$ \cite{vogel87}, H$_2$CO \cite{martin-pintado90}, CH$_3$CN
\cite{devicente97} and 
C$_2$H$_3$CN \cite{liu99}.
Different molecules seem to give different physical properties and kinematics
for the hot cores. This is because different tracers require different 
excitation conditions and because of the complexity of the region.
Cyanoacetylene (HC$_3$N) is a very suitable molecule to probe the 
physical properties and kinematics in the hot cores since it has several 
vibrational states which lie in the range between 300 and 1000 K, which 
are only excited by IR radiation in the hot cores. Thus the vibrationally 
excited lines of HC$_3$N probe regions with strong emission in the mid-IR.

Since the detection of the rotational line from the vibrational
v7=1 state by Goldsmith et al. \cite*{goldsmith82}
in Orion, rotational lines from vibrational states arising from other 
sources and from
other vibrational states have been discovered. 
Most of the studies of vibrationally excited HC$_3$N 
(hereafter HC$_3$N*) emission towards Sgr B2 have been made with low 
angular resolution 
\cite{goldsmith85,martin-pintado86} and only towards Sgr B2M and Sgr B2N.

Recently 
Mart\'\i n-Pintado et al. \cite*{martin-pintado99} have found that 
recent massive star formation
in the stage of hot cores is not only restricted to Sgr B2M, Sgr B2N, 
and Sgr B2S but is also present in the envelope of Sgr B2. These new hot cores
are less massive than those in Sgr B2M and Sgr B2N.  This suggests
that recent massive star formation similar to that in Orion A might occur 
over more
extended regions than the main cores Sgr B2M, B2N and B2S.
 The aim of this paper is twofold: to obtain a global picture of massive star 
formation between the main cores Sgr B2M and B2N and to study in 
detail the hot cores Sgr B2M and Sgr B2N. To study
in detail the physical properties and the 
kinematics of the hot cores in Sgr B2M and Sgr B2N, we present 
interferometric observations of several HC$_3$N* lines.
The global picture of star formation has been obtained from mapping several 
HC$_3$N* lines obtained with the IRAM 30m dish. This map has revealed for 
the first time the presence of a ridge of hot cores similar to that
in Orion A.

\section{Observations}

\subsection{Single dish observations}

We have carried out single dish observations of different rotational lines 
of HC$_3$N and its isotopic species HC$^{13}$CCN and HCC$^{13}$CN from 
the ground and vibrational states with the IRAM 30m 
telescope at Pico de Veleta (Spain) towards the regions containing 
the hot cores Sgr B2M and 
Sgr B2N. We observed the J=10--9 (3.3 mm), J=12--11 (2.9 mm), 
J=15--14 (2.2 mm), J=16--15 (2 mm), J=17--16 (1.9 mm) and J=24--23 (1.3 mm). 
The Half Power Beam Width (HPBW) of the telescope at 3, 2 and 1.3 mm 
was 26$''$, 17$''$ and 12$''$. 
We used three SIS receivers to observe simultaneously the 3, 2 and 1.3 mm 
lines. 
Most of the observations were made with the receivers tuned to SSB mode with
an image rejection of 7 to 9 dB. The observations were performed in a 
position 
switching mode, taking the reference position 10$'$ off the on-sources 
in right ascention. The line intensity calibration was established 
measuring a cold and ambient temperature load of known temperatures. The 
line intensities have 
been converted to main beam temperatures by using beam efficiencies
of 0.6 for 3 and 2mm and 0.45 for 1.3 mm and a forward eficciency 
of 0.9 for all wavelengths. As spectrometers we used two 512 MHz filter 
banks
and one 1024 channel autocorrelator. The velocity resolution provided by the
spectrometers were 3.3, 2.7, 1.9 and 0.4 kms$^{-1}$ for the 3.3, 2.9, 2 and 
1.3 mm lines respectively. 

In order to determine the spatial extension of the HC$_3$N* 
emission, we mapped a region of 50$''\times$ 50$''$ 
around Sgr B2M and Sgr B2N in the J=16--15,3v7,1e and the 
J=24--23,3v7,1e ${\rm HC_3N}$ lines with a 5$''$ spacing. Simultaneous 
with the J=16,1e-15,3v7,1e line we observed the ${\rm H_2^{13}CO}$ 
2(1,1)--2(1,0)
transition which is seen in absorption towards both continuum sources and 
allowed to 
check the pointing relative to the continuum sources in our map. 
Observations towards the most intense condensations in the previous map were
also made in the J=10-9,1v7,1f 1v6,1e and 1v5,1f transitions and in the
J=24-23,1v7,1f transition.

\subsection{Interferometric observations}
Since the single dish HC$_3$N* maps 
showed this emission to be unresolved towards the 
Sgr B2M and Sgr B2N cores we used the IRAM interferometer 
at Plateau de Bure, France, to map the emission with higher angular 
resolution. We performed two snapshots of Sgr B2M 
and Sgr B2N in July 1995 with three antennas. The instrumental parameters 
for the Plateau de Bure interferometer observations are summarized 
in Table \ref{t:pdbI}. Observations were made at 91.2 GHz in order 
to measure simultaneously the 
J=10--9,1v5,1f, J=10--9,1v6,1e, J=10--9,1v7,1e and J=10--9,1v7,1f transitions.
We used the source 1730-13 (14 Jy) as the flux density and bandpass 
calibrator and 1830-21 (3.2 Jy) as the phase calibrator. 

\begin{table}
 \caption{Instrumental parameters for the PdBI observations} 
 \begin{flushleft}                                                              
\begin{tabular}{lr}                                                   
\multicolumn{1}{l}{Parameter}     & \multicolumn{1}{c}{Value} \\ \hline
Sources observed                  & 2       \\  
Total time on sources             & 1.6 hours       \\  
Central sky frequency             & 91.155 GHz \\
Primary beam (HPBW)               &         56$''$ \\
Synthesized beam (HPBW) (B2M)     & $10.5''\times 4.9 ''$ PA: 0$^o$ \\
Synthesized beam (HPBW) (B2N)     & $12.2''\times 4.4 ''$ PA: $-14^o$ \\
Interferometer configuration      & Dc (W05,E03,N05) \\
Flux density calibrator           & 1730-130 (14 Jy) \\
Phase and Band Pass calibrators   & 1730-130 (14 Jy) \\
                                  & 1830-21  (3.2 Jy) \\
Phase center position J2000 (B2M) & 17:47:20.3 -28:23:07 \\
Phase center position J2000 (B2N) & 17:47:20.3 -28:22:19 \\
Absolute position error           & $\pm$  $0.4''$ \\ \hline
\end{tabular}                                                                   
\end{flushleft}
\label{t:pdbI}
\end{table}

\section{Results}
\subsection{Single dish}

\begin{figure}
\resizebox{\hsize}{!}{\includegraphics{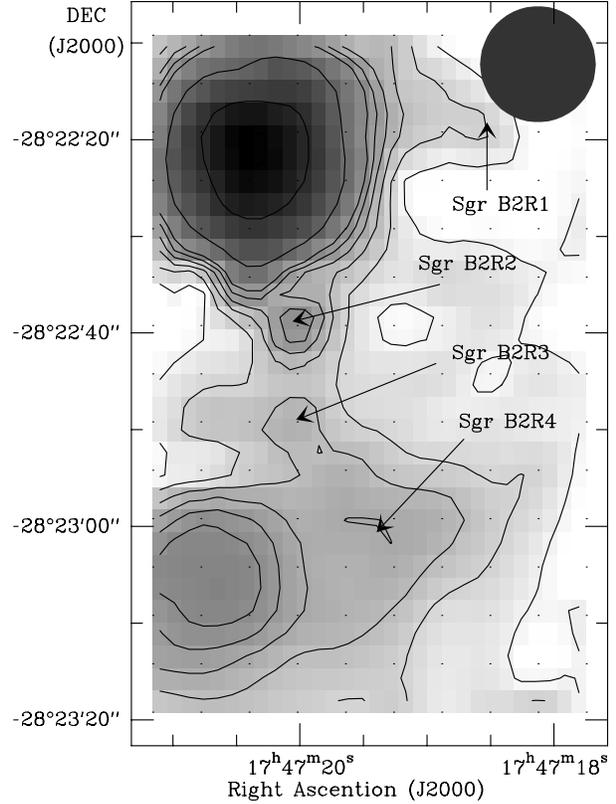}}
\caption{Map obtained at the 30m radio telescope for the integrated intensity 
of HC$_3$N 
J=24-23,3v7,1e transition (see Sect. 3.1 for an additional explanation). 
Levels are 5, 10 15, 20, 25, 30, 50, 100, 
and 150 K\, kms$^{-1}$ using antenna temperature scale. We have labeled 
the new hot cores along the ridge as Sgr B2R1, B2R2, B2R3 and B2R4. The filled
circle in the upper right corner shows the beam size.}
\label{f:30m_map}
\end{figure}

 Fig. \ref{f:30m_map} shows a map of the integrated intensity of the 
J=24--23,3v7,1e ${\rm HC_3N}$ line.
This line was observed with the receiver tuned in nearly double side 
band, the relative
gain for the image band was only 0.2.
The integrated emission also contains the J=25--24,1v7,1e and 1v6,1e lines which
were observed in the image band. The map shows the morphology
of the HC$_3$N$^*$ in the region. The HC3N* emission is dominated by 
the already known massive hot cores associated 
with Sgr B2M and Sgr B2N 
\cite{goldsmith87,vogel87,lis91,martin-pintado90}. 
The map also shows for the first time
the presence of a ridge of HC$_3$N*  emission
connecting the Sgr B2M and B2N hot cores, and an extended low
brightness emission to the west of the ridge. The HC$_3$N*
ridge is unresolved in the east-west direction. For further
study we have considered four sources labelled as
B2R1, B2R2, B2R3 and B2R4 (see Fig. \ref{f:30m_map}). The extended 
component has a
size of $\sim 30"$, and hereafter will be referred to as Sgr B2W.

\begin{figure}
\resizebox{\hsize}{!}{\includegraphics{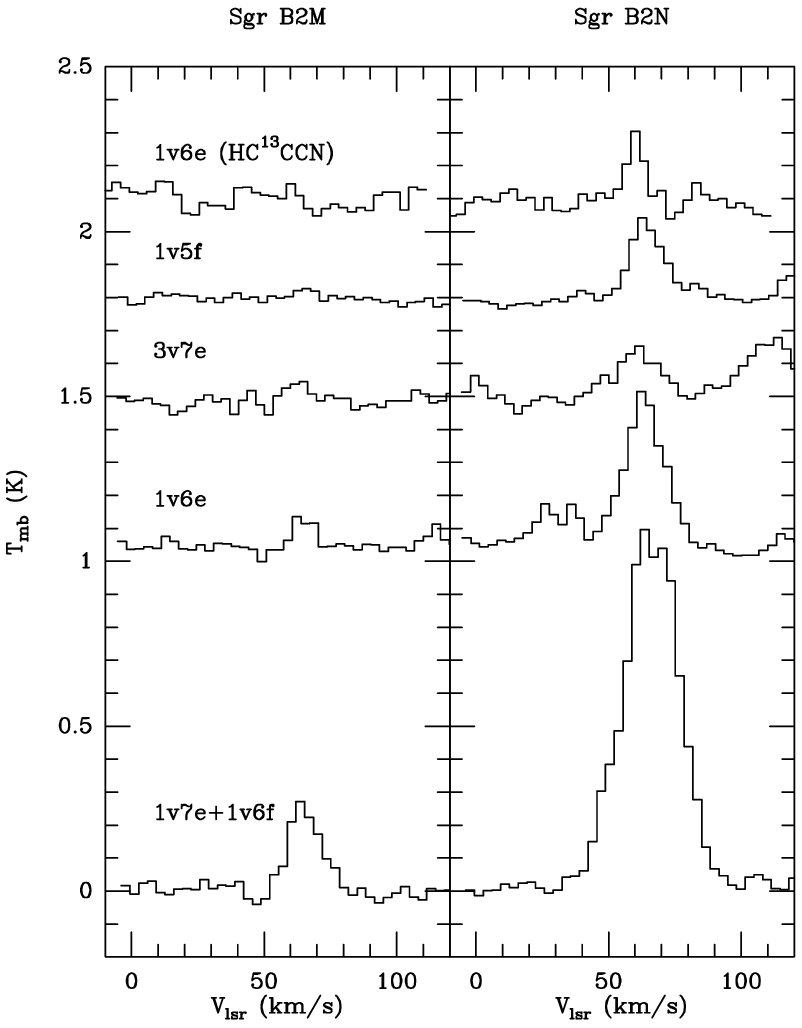}}
 \caption{Spectra taken at the 30m radio telescope for the HC$_3$N J=10-9 
          transition in different vibrational excited levels, 
          towards Sgr B2M and Sgr B2N. The upper panel shows the spectra
          of the J=10-9 line of HC$^{13}$CCN in the 1v6,1e vibrational level.}
 \label{f:30m_spec}
\end{figure}

Fig. \ref{f:30m_spec} shows typical profiles of the J=10--9 line of 
HC$_3$N* and 
HC$^{13}$CCN for several vibrational states towards Sgr B2M (left panel) and
Sgr B2N (right panel) taken at the 30m radio telescope. This figure 
clearly shows the difference in the 
excitation of the material in the Sgr B2M and Sgr B2N hot cores.
Table \ref{t:sdish_res} summarizes the single dish results. The line 
parameters in this Table have been derived by fitting a Gaussian 
profile to the lines.
As intensity scale we have used the main beam temperature because 
the emission region is in all cases smaller than the telescope beam. 

The rotational transitions of the vibrational states appear in
pairs, due to the level degeneracy, separated in frequency by tenths
of MHz. This feature has been used to aid in the determination of the
parameters of the Gaussian fits when one of the transition pairs appears
contaminated by more intense lines. In those cases in which we have used 
the width of the pair transition, the error associated with the fitted 
parameter of the contaminated line is indicated in Table \ref{t:sdish_res} 
as 0. The line widths towards Sgr B2N and Sgr B2M depend on the energy of 
levels involved in the transition. The line width increases as the energy of 
the levels involved increases.

\begin{table*}
 \caption{Fits for different ${\rm HC_3N}$ transitions towards Sgr B2M 
and Sgr B2N. Single dish observations}
 \begin{flushleft}
 \begin{tabular}{lrrrrrrrrr}
 \multicolumn{2}{l}{} & \multicolumn{4}{c}{Sgr B2M} & 
 \multicolumn{4}{c}{Sgr B2N} \\ \hline
Transition & ${\rm E_{upp}}$ & $T_{mb}$& $v$ & $\Delta v$ & Area &
                               $T_{mb}$& $v$ & $\Delta v$ & Area \\
             & (K)  & (K) &(${\rm km\,s^{-1}}$) & (${\rm km\,s^{-1}}$) & 
               (${\rm K\, km\,s^{-1}}$)
                    & (K) &(${\rm km\,s^{-1}}$) & (${\rm km\,s^{-1}}$) & 
               (${\rm K\, km\,s^{-1}}$) \\ \hline 

10--9            &24.1 &7.45&59.2(0.1)&19.3(0.3)&153.0(1.9)& 
                        7.24&66.5(0.1)&23.6(0.1)&182.0(0.7) \\ \hline
10,1e--9,1e\ 1v5 &978.3&    &         &         &         & 
                        0.43&62.3(0.6)&13.9(1.0)& 6.4(0.5) \\ \hline
10,1f--9,1f\ 1v5 &978.3&0.04&65.5(1.3)& 7.3(2.5)& 0.3(0.1)&
                        0.32&63.9(0.8)&15.3(1.9)& 5.2(0.6)\\ \hline
10,1e--9,1e\ 1v6 &741.3&0.09&64.8(0.6)&11.3(1.4)& 1.0(0.1)& 
                        0.54&63.6(0.7)&17.6(1.7)&10.1(0.8)\\ \hline
10,1f--9,1f\ 1v6 &741.3&0.07&62.8(0.0)& 8.4(0.0)& 0.7(0.0)& 
                        0.49&63.6(0.1)&16.1(1.1)& 8.3(0.5)\\
10,1e--9,1e\ 1v7 &344.1&0.36&64.1(0.3)&11.7(0.8)& 4.4(0.3)& 
                        1.13&63.0(0.1)&23.0(0.6)&27.5(0.7)\\ \hline
10,0--9,0 \ 2v7  &662.1&    &         &         &         &
                        0.22&64.0(1.0)&24.2(1.3)& 5.6(0.4)\\
10,2e--9,2e \ 2v7 &664.7&    &         &         &         &
                        0.22&63.9(0.0)&24.2(0.0)& 5.6(0.0)\\
10,2f--9,2f \ 2v7 &664.7&    &         &         &         &
                        0.22&64.0(0.0)&24.2(0.0)& 5.6(0.0)\\ \hline
10,1e--9,1e\ 3v7 &980.6&0.08&62.6(1.2)& 8.9(2.2)& 0.7(0.2)& 
                        0.23&61.1(1.4)&14.0(0.0)& 3.5(0.4)\\ \hline
12--11           &34.1 &9.19&59.4(0.1)&18.1(0.3)&177.1(2.3)& 
                        8.33&66.0(0.1)&22.1(0.2)&196.3(1.3) \\ \hline
12,1e--11,1e\ 1v5 &988.3&   &         &         &         &
                        0.78&63.4(0.7)&15.8(1.4)&13.1(1.1)\\ \hline
12,1f--11,1f\ 1v5 &988.3&   &         &         &         &
                        0.75&63.0(0.0)&12.0(0.0)& 9.6(0.0)\\ \hline
12,1e--11,1e\ 1v6&751.3&0.16&62.9(1.0)&11.6(2.5)& 2.0(0.3)&
                        0.60&62.4(0.5)&19.2(1.2)&12.1(0.6)\\ \hline
12,1f--11,1f\ 1v6&751.4&0.15&63.0(0.0)&11.6(0.0)& 1.9(0.0)&
                        0.90&64.0(0.0)&15.8(0.0)&15.1(1.6)\\
12,1e--11,1e\ 1v7&354.1&0.68&62.7(0.0)&17.0(0.0)&12.3(1.1)&
                        2.11&62.2(0.0)&26.0(0.0)&58.5(0.0)\\ \hline
12,1f--11,1f\ 1v7&354.1&0.87&61.1(2.0)&17.0(0.0)&15.7(2.1)& 
                        1.75&62.2(1.4)&28.2(3.4)&60.8(4.5)\\ \hline
12,0--11,0 \ 2v7  &672.2&   &         &         &         &
                        0.80&62.9(2.7)&17.2(2.7)&13.8(2.1)\\
12,1e--11,1e\ 2v7 &672.2&   &         &         &         &
                        0.80&62.9(0.0)&13.8(0.0)&17.2(0.0)\\
12,1f--11,1f\ 2v7 &672.2&   &         &         &         &
                        0.80&62.9(0.0)&13.8(0.0)&17.2(0.0)\\ \hline
15--14            &52.4 &7.31&61.4(0.1)&17.0(0.3)&131.9(2.3)& 
                         7.62&64.5(0.3)&20.4(0.9)&165.4(5.5) \\ \hline
15,1e--14,1e\ 1v5 &1006.7&  &         &         &         &
                        0.88&63.0(0.0)&14.1(0.0)&15.0(0.0)\\ \hline
15,1f--14,1f\ 1v5&1006.7&0.14&63.0(0.0)&12.0(0.0)&1.7(0.0)&
                        0.48&63.0(0.0)& 15.0(0.0)&7.7(0.0)\\ \hline
16--15            &59.4 &7.73&60.3(0.1)&17.8(0.2)&146.2(1.5)& 
                         8.84&65.7(0.3)&22.1(0.8)&206.4(5.8) \\ \hline
16,1e--15,1e\ 3v7&1016.3&0.22&61.9(1.7)&10.0(0.0)&2.4(0.4)&
                        0.65&65.5(1.6)&14.2(3.6)&9.87(2.1)\\ \hline
17--16            &66.8 &17.75&62.3(0.1)&14.9(0.2)&282.4(3.5)& 
                         11.20&64.2(0.1)&20.0(0.7)&240.1(6.2) \\ \hline
17,1f--16,1f\ 1v5 &1021.1&  &         &         &         &
                        1.24&63.3(0.4)&10.0(0.0)&13.2(0.7)\\ \hline
17,1f--16,1f\ 1v7&387.1&2.19&62.3(0.3)&15.7(0.6)&36.5(1.2)&
                        3.99&64.7(0.3)&25.5(0.8)&108.3(3.0)\\ \hline
17,0--16,0  \ 2v7&705.2&0.76&62.9(0.0)&12.0(0.0)& 9.7(0.8)&
                        1.56&71.9(0.6)&24.1(0.8)&40.1(0.8)\\
17,2e--16,2e\ 2v7&707.8&0.76&62.9(0.0)&12.0(0.0)& 9.7(0.0)&
                        1.56&71.9(0.0)&24.1(0.8)&40.1(0.0)\\
17,2f--16,2f\ 2v7&707.8&0.76&62.9(0.0)&12.0(0.0)& 9.7(0.0)&
                        1.56&71.9(0.0)&24.1(0.8)&40.1(0.0)\\ \hline
17,1e--16,1e\ 3v7 &1023.7&  &         &         &         &
                        1.0 &62.6(1.4)&11.1(2.5)&11.8(2.7)\\ \hline
24,1e--23,1e\ 1v6&848.6&0.53&63.5(0.0)&11.0(0.0)& 6.3(2.2)&
                        1.59&78.7(2.1)&14.4(4.7)&24.3(0.8)\\ \hline
24,1e--23,1e\ 1v7&451.3&2.06&63.1(0.8)&17.0(0.0)&37.3(1.7)&
                        5.89&66.9(0.8)&23.6(2.6)&147.8(0.8)\\ \hline
24,1f--23,1f\ 1v7&451.5&1.9 &62.6(0.4)&18.8(0.8)&38.7(1.5)&
                        5.54&65.9(0.1)&26.5(1.8)&156.6(0.9)\\
\hline
\end{tabular}
\end{flushleft} 
\label{t:sdish_res}
\end{table*}

Table \ref{t:sdish_iso} summarizes the Gaussian fits to the 
${\rm HC^{13}CCN}$ and ${\rm HCC^{13}CN}$ single dish profiles at the
ground and vibrational excited levels. 
The frequencies of the HC$^{13}$CCN and HCC$^{13}$CN lines have been calculated
using the expressions given by Wyrowski et al. \cite*{wyrowski99}.
Due to line confusion we have only used those transitions 
which were not blended with other lines.
The rotational transitions for the vibrational ground state of HC$^{13}$CCN and 
HCC$^{13}$CN are separated a few MHz and can be observed simultaneously. Both 
lines have similar intensities and widths indicating that both 
isotopic substitutions have the same abundance, within the calibration errors.

From the intensities of the main isotope 
and from those of the $^{13}$C isotope, and assuming an isotopic 
ratio $^{12}$C/$^{13}$C of 20 for the Galactic 
center \cite{wilson94}, we estimate the opacity depth of the main 
isotope HC$_3$N lines. Towards Sgr B2M the opacities of the main isotope 
rotational transitions in the ground vibrational state 
are between 0.4 and 0.6, while 
for the vibrational levels we obtain an upper limit of 0.1.
Therefore all HC$_3$N$^*$ lines in the Sgr B2M hot core are optically thin.
On the other hand the opacity of the HC$_3$N* lines towards Sgr B2N 
is $\sim$ 4.5. This means that HC$^{13}$CCN and HCC$^{13}$CN lines in 
the ground states are also probably optically thick and cannot be used to 
determine the opacity of the main isotope transitions.

%
%
%
%

\begin{table*}
 \caption{Fits for different ${\rm HC^{13}CCN}$ (a) and ${\rm HCC^{13}CN}$ (b) 
transitions towards Sgr B2M and Sgr B2N. Single dish observations}
 \begin{flushleft}
 \begin{tabular}{lrrrrrrrrr}
 \multicolumn{2}{l}{} & \multicolumn{4}{c}{Sgr B2M} & 
 \multicolumn{4}{c}{Sgr B2N} \\ \hline
Transition & ${\rm E_{upp}}$ & $T_{mb}$& $v$ & $\Delta v$ & Area &
                               $T_{mb}$& $v$ & $\Delta v$ & Area \\
             & (K)  & (K) &(${\rm km\,s^{-1}}$) & (${\rm km\,s^{-1}}$) & 
               (${\rm K\, km\,s^{-1}}$)
                    & (K) &(${\rm km\,s^{-1}}$) & (${\rm km\,s^{-1}}$) & 
               (${\rm K\, km\,s^{-1}}$) \\ \hline 
10,1e--9,1e\ 1v6 &741.6&$<$0.01    &         &         &         & 
                        0.19&60.1(0.4)&7.0(0.9)& 1.5(0.1) \\ \hline
10,1f--9,1f\ 1v7 &343.9&$<$0.03    &         &         &         &
                        0.25&63.5(1.4)&17.3(2.6)& 4.6(0.4)\\ \hline
10,1e--9,1e\ 2v7 &664.0&$<$0.01    &         &         &         & 
                        0.12&64.9(1.2)&17.7(3.4)& 2.4(0.2)\\ \hline
12,1f--11,1f\ 1v7 &354.0&$<$0.07   &         &         &         &
                        0.21&62.9(1.5)&17.5(4.0)& 3.9(0.5)\\ \hline
15--14            &52.2 &0.46&62.2(0.7)&15.1(1.8)&5.3(0.5)& 
                         1.13&65.5(1.2)&22.4(2.7)&26.9(3.6) \\
15--14            &52.2 &0.53&62.2(0.6)&13.8(1.3)&5.5(0.4)& 
                         0.98&68.1(1.6)&24.3(3.3)&25.3(3.6) \\ \hline
16--15            &59.2 &0.68&64.0(0.4)&13.5(0.9)&9.6(0.6)& 
                         1.03&67.5(0.6)&26.1(1.3)&28.7(1.4) \\
16--15            &59.2 &0.67&62.6(0.4)&13.4(0.9)&9.6(0.6)& 
                         1.10&65.6(0.4)&18.3(1.0)&21.4(1.2) \\ \hline
17--16            &66.5 &0.56&34.1(0.5)&17.6(1.6)&10.4(0.7)& 
                         1.14&65.9(1.3)&22.0(0.0)&26.6(1.2) \\
17--16            &66.5 &0.51&63.5(0.4)&15.2(1.0)&8.2(0.5)& 
                         0.98&63.4(0.9)&23.4(1.8)&24.6(1.8) \\ \hline
17,1f--16,1f\ 1v7 &384.9&$<$0.07   &         &         &         &
                        1.03&65.3(0.5)&16.9(1.1)& 18.7(0.7)\\ \hline

\end{tabular}
\end{flushleft} 
\label{t:sdish_iso}
\end{table*}

Table \ref{t:other_cores} lists the results of the Gaussian fits towards 
the new hot cores 
Sgr B2R1, Sgr B2R2, Sgr B2R3 and Sgr B2R4 determined from the 30m dish 
observations of the 10--9, 1v7,1f, 1v6,1e and 1v5,1f and 24--23,1v7,1f 
transitions. No $^{13}$C isotope lines of HC$_3$N have been measured for 
these sources.

\begin{table}
 \caption{Fits for different ${\rm HC_3N}$ vibrationally excited transitions 
towards Sgr B2R1, Sgr B2R2, Sgr B2R3 and Sgr B2R4. Single dish observations}
 \begin{flushleft}
 \begin{tabular}{lrrrr}
 \multicolumn{5}{c}{Sgr B2R1} \\ \hline
Transition &  $T_{mb}$& $v$ & $\Delta v$ & Area \\
             & (K)  & (K) &(${\rm km\,s^{-1}}$) & 
             (${\rm K\, km\,s^{-1}}$) \\ \hline
10,1f--9,1f\ 1v5 &0.06&67.4(2.0)&12.0(0.0)& 0.84(0.3) \\ \hline
10,1e--9,1e\ 1v6 &0.20&63.8(0.7)&11.7(1.7)& 2.45(0.4) \\ \hline
10,1e--9,1f\ 1v7 &0.43&65.5(0.4)&23.7(1.0)&11.04(0.5) \\ \hline
24,1f--23,1f\ 1v7&0.40&57.4(0.8)& 7.5(2.3)& 3.20(1.0) \\ \hline
 \multicolumn{5}{c}{Sgr B2R2} \\ \hline
10,1f--9,1f\ 1v5 &0.02&64.8(2.9)&17.0(0.0)& 0.45(0.2) \\ \hline
10,1e--9,1e\ 1v6 &0.10&63.9(1.0)&17.7(2.7)& 1.90(0.3) \\ \hline
10,1f--9,1f\ 1v7 &0.24&63.2(0.5)&19.1(1.5)& 4.91(0.4) \\ \hline
24,1f--23,1f\ 1v7&0.28&59.6(0.4)& 9.0(1.0)& 2.70(0.2) \\ \hline
 \multicolumn{5}{c}{Sgr B2R3} \\ \hline
10,1e--9,1e\ 1v6 &0.11&65.3(1.1)& 8.9(2.7)& 1.07(0.4) \\ \hline
10,1f--9,1f\ 1v7 &0.26&63.4(1.0)&14.0(0.0)& 3.86(0.5) \\ \hline
24,1f--23,1f\ 1v7&0.24&62.4(0.7)&13.1(1.6)& 5.01(0.5) \\ \hline
 \multicolumn{5}{c}{Sgr B2R4} \\ \hline
10,1e--9,1e\ 1v6 &0.12&65.3(0.8)& 8.0(0.0)& 1.03(0.2) \\ \hline
10,1f--9,1f\ 1v7 &0.29&63.1(0.6)&17.2(1.9)& 5.30(0.5) \\ \hline
24,1f--23,1f\ 1v7&0.34&58.8(0.8)&16.9(1.6)& 6.06(0.6) \\ \hline
\end{tabular}
\end{flushleft} 
\label{t:other_cores}
\end{table}

\subsection{Interferometric data}

\begin{figure*}
\rotatebox{-90}{\includegraphics{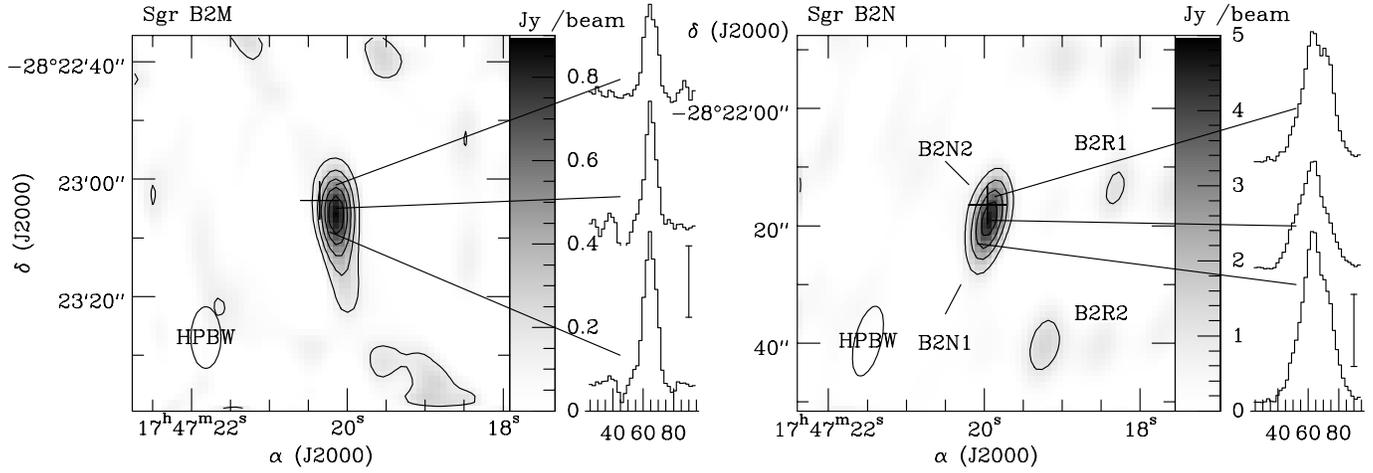}}
 \caption{Line emission from transition HC$_3$N* 10-9,1v7,1f towards Sgr B2M
  and Sgr B2N as observed with the Plateau de Bure Interferometer.
  The cross represents the continuum position at 91 GHz. Sgr B2M: levels 
  at 0.14, 0.29, 0.43, 0.58, 0.72 and 0.87 Jy/beam. Sgr B2N: 
  levels at 1.2, 2.3, 3.4 and 4.5 Jy/beam. The vertical bar accompanying 
  the spectra represents an intensity of 0.5 Jy/beam 
  in Sgr B2M and 2 Jy/beam in Sgr B2N.}
 \label{f:map_pdb}
\end{figure*}

\begin{figure*}
\rotatebox{-90}{\includegraphics{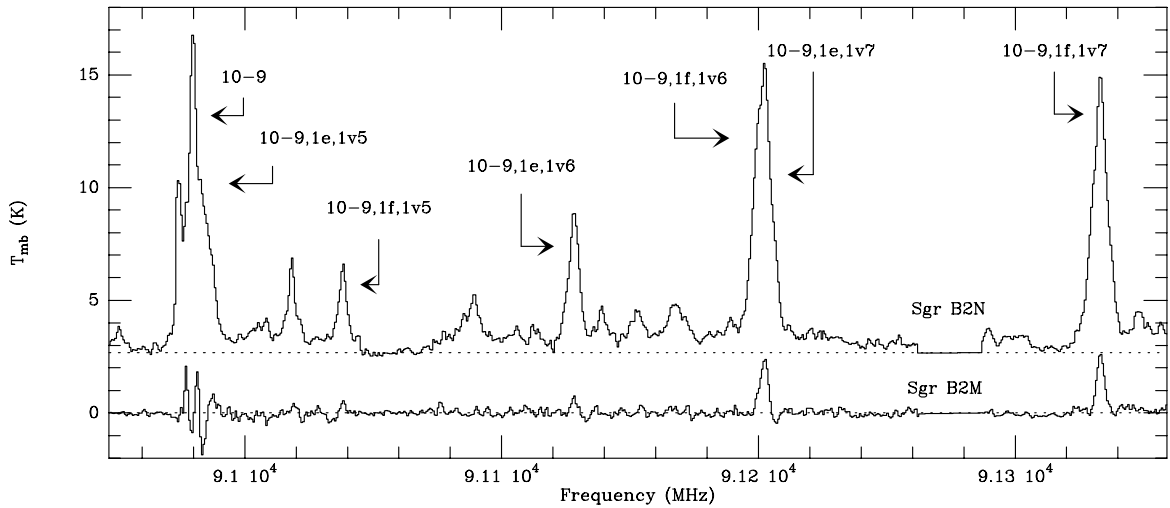}}
 \caption{Spectra of the HC$_3$N* lines towards Sgr B2M (lower) and Sgr B2N
 (upper) as obtained with the Plateau de Bure Interferometer.} 
 \label{f:pdb_spec}
\end{figure*}

Figs. \ref{f:map_pdb}a and \ref{f:map_pdb}b show the maps 
of the integrated intensity of the HC$_3$N J=10--9,1v7,1f line 
(at 64 kms$^{-1}$) towards 
Sgr B2M and Sgr B2N obtained with the PdB interferometer. The 
crosses show the location of the continuum peaks measured with the PdBI 
at 3 mm in both regions. The continuum flux densities are 3 Jy/beam 
for Sgr B2M and 2 Jy/beam for Sgr B2N. The HC$_3$N$^*$ line emission 
is unresolved towards both hot cores with an upper limit to the size 
of the emitting regions of $3''\times 5''$ (1/2 the HPBW). 
The peak of the HC$_3$N* emission is shifted from the continuum 
by $\sim$ 3$''$ to the 
southeast for Sgr B2M and to the south for Sgr B2N.
The condensations Sgr B2R1 and Sgr B2R2 are also seen at the border of 
the interferometric maps, although close to the noise limit
(Fig. \ref{f:map_pdb}b). 
Fig. \ref{f:pdb_spec} shows the spectra taken towards the maximum 
of the HC$_3$N* emission in Sgr B2M (lower) and Sgr B2N (upper). 
We have used the main beam temperature as the intensity scale (conversion
factor of 2.9 Jy/K). 

Table \ref{t:pdb_res} summarizes 
the results of the Gaussian fits performed to the HC$_3$N* profiles towards 
Sgr B2M and Sgr B2N. Lines towards Sgr B2N show two velocity components 
at 63 and 75 kms$^{-1}$. Each velocity corresponds to a different hot core, as
we will see in Sect. 4.
The line fluxes measured with the interferometer towards Sgr B2M 
and Sgr B2N are within the calibration errors (20\%) 
of those obtained from single dish. Therefore,
as expected, all the emission from HC$_3$N$^*$
arises from compact sources unresolved with our resolution.

\begin{table*}
 \caption{Fits for different ${\rm HC_3N}$ vibrationally excited transitions 
towards Sgr B2M and Sgr B2N. Interferometric observations}
 \begin{flushleft}
 \begin{tabular}{lrrrrrrrrr}
 \multicolumn{2}{l}{} & \multicolumn{4}{c}{Sgr B2M} & 
 \multicolumn{4}{c}{Sgr B2N} \\ \hline
Transition & ${\rm E_{upp}}$ & $T_{mb}$& $v$ & $\Delta v$ & Area &
                               $T_{mb}$& $v$ & $\Delta v$ & Area \\
             & (K)  & (K) &(${\rm km\,s^{-1}}$) & (${\rm km\,s^{-1}}$) & 
               (${\rm K\, km\,s^{-1}}$)
                    & (K) &(${\rm km\,s^{-1}}$) & (${\rm km\,s^{-1}}$) & 
               (${\rm K\, km\,s^{-1}}$) \\ \hline 
10,1f--9,1f\ 1v5 &978.3&0.52&65.2(0.4)& 6.6(1.2)& 3.7(0.6)&
                        3.55&63.9(0.1)&10.7(0.1)&40.6(0.5)\\ 
                 &     &    &         &         &          &
                        0.41&75.0(0.0)& 5.0(0.8)& 2.2(0.4)\\ \hline
10,1e--9,1e\ 1v6 &741.3&0.73&64.9(0.6)& 8.2(1.2)& 6.4(0.9)&
                        6.12&62.8(0.2)&14.9(0.5)&97.5(2.8)\\ 
                 &     &    &         &         &          &
                        1.11&75.6(0.6)& 6.4(0.9)& 7.6(2.1)\\ \hline
10,1f--9,1f\ 1v6 &741.3&0.69&65.4(0.3)& 6.8(1.6)& 5.2(0.9)& 
                        1.60&69.3(0.9)&16.0(0.0)&27.0(1.5)\\
10,1e--9,1e\ 1v7 &344.1&2.51&65.4(0.3)&10.3(0.0)& 27.4(0.9)& 
                        10.77&64.6(0.2)&23.0(0.0)&262.7(1.4)\\ \hline
10,1e--9,1f\ 1v7 &344.1&2.48&65.1(0.2)&10.9(0.4)& 28.8(0.9)&
                       11.91&62.2(0.2)&17.0(0.0)&215.6(4.4)\\ 
                 &     &    &         &         &          &
                        4.51&75.0(0.0)&15.2(1.1)& 73.0(5.8)\\  \hline
\end{tabular}
\end{flushleft} 
\label{t:pdb_res}
\end{table*}

%
%
%


\section{Velocity structure in the Sgr B2M and B2N cores}

The gaussian fits obtained from single dish and interferometric observations
show that the rotational lines 
arising from different vibrational levels have different line widths. The
linewidth decreases as the energy of the vibrational level increases. In 
general,
the 1v6 lines are narrower than the 1v7 lines by 
nearly a factor of 2 (see Table \ref{t:l_width} for a summary). For 
Sgr B2N the lines are optically thick and opacity could explain some 
of the line 
broadening. However the H$^{13}$CCCN lines in the 1v7 and 1v6 vibrationally
excited states which are optically thin also show the same trend.

\begin{table}
 \caption{Line widths averaged for different ${\rm HC_3N}$ vibrationally 
 excited transitions towards Sgr B2M and Sgr B2N. Single dish observations}
 \begin{flushleft}
 \begin{tabular}{lrrrr}
 \multicolumn{1}{l}{} & \multicolumn{1}{c}{Sgr B2M} & 
 \multicolumn{2}{c}{Sgr B2N} \\ \hline
 \multicolumn{1}{l}{} & \multicolumn{1}{c}{$^{12}$C} & 
 \multicolumn{1}{c}{$^{13}$C} & \multicolumn{1}{c}{$^{12}$C} &
 \multicolumn{1}{c}{$^{13}$C} \\
Transition & $\Delta v$ (km/s) & $\Delta v$ (km/s) & 
             $\Delta v$ (km/s) & $\Delta v$ (km/s) \\ \hline
v0   &17.4(0.8)&14.3(0.4)&21.6(0.7)& 21.8(1.7)\\
v7   &16.2(1.0)&         &25.5(1.1)& 17.0(1.0)\\
2v7  &12.0(1.0)&         &21.8(2.8)& 17.7(3.4)\\
v6   &10.8(0.7)&         &16.6(0.9)&  7.0(0.9)\\
3v7  & 9.5(0.5)&         &13.3(0.9)&         \\
v5   & 9.6(2.3)&         &13.8(0.9)&         \\
\hline
\end{tabular}
\end{flushleft} 
\label{t:l_width}
\end{table}

\begin{figure*}
\rotatebox{-90}{\includegraphics{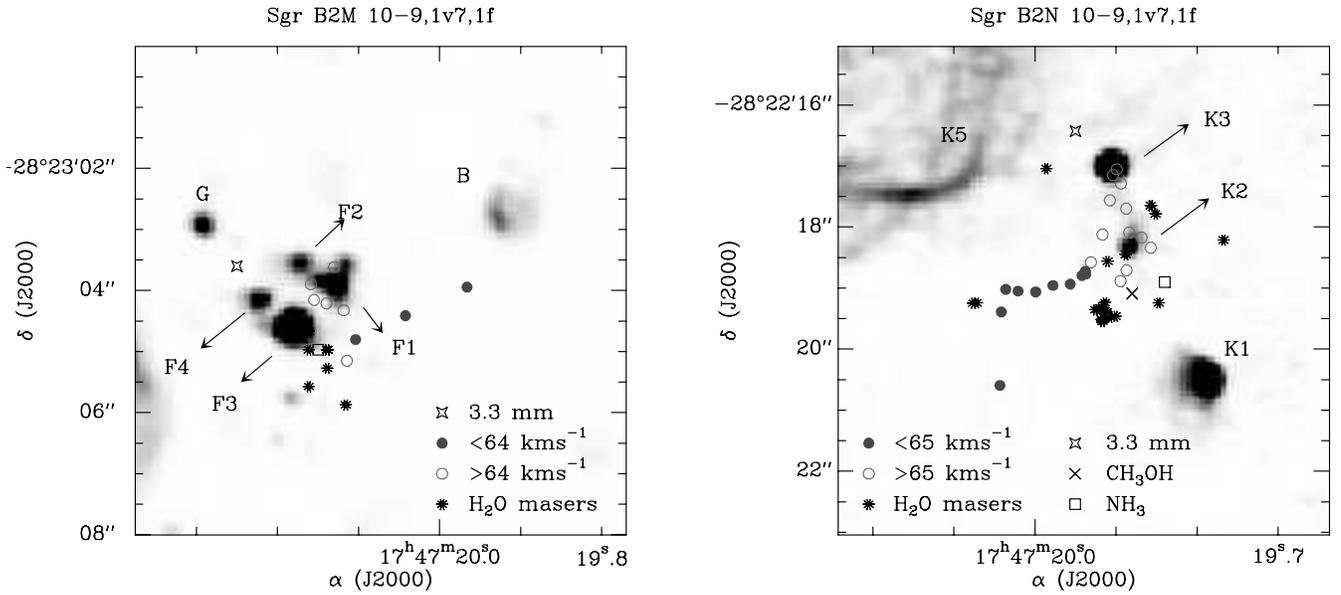}}
 \caption{Location of the HC$_3$N* 10--9,1v7,1f velocity features (circles)
  towards Sgr B2N and Sgr B2M, as obtained from the interferometer using an UV 
  fitting technique. Circles are plotted in steps of 2.1 kms$^{-1}$. The 
  background image is the 1.3 cm continuum image from 
  Gaume et al. (1995). Asterisks correspond to H$_2$O maser 
  positions from Reid et al. (1988) (Sgr B2N) and Kobayashi et al.
  (1989) (Sgr B2M), diamonds 
  to 3.3 mm peak emission (this article), squares to NH$_3$ thermal cores 
  (Vogel et al. (1987), and crosses to CH$_3$OH thermal cores (Mehringer \& Menten, 1997) 
  }
 \label{f:spots}
\end{figure*}

\begin{figure}
\resizebox{\hsize}{!}{\includegraphics{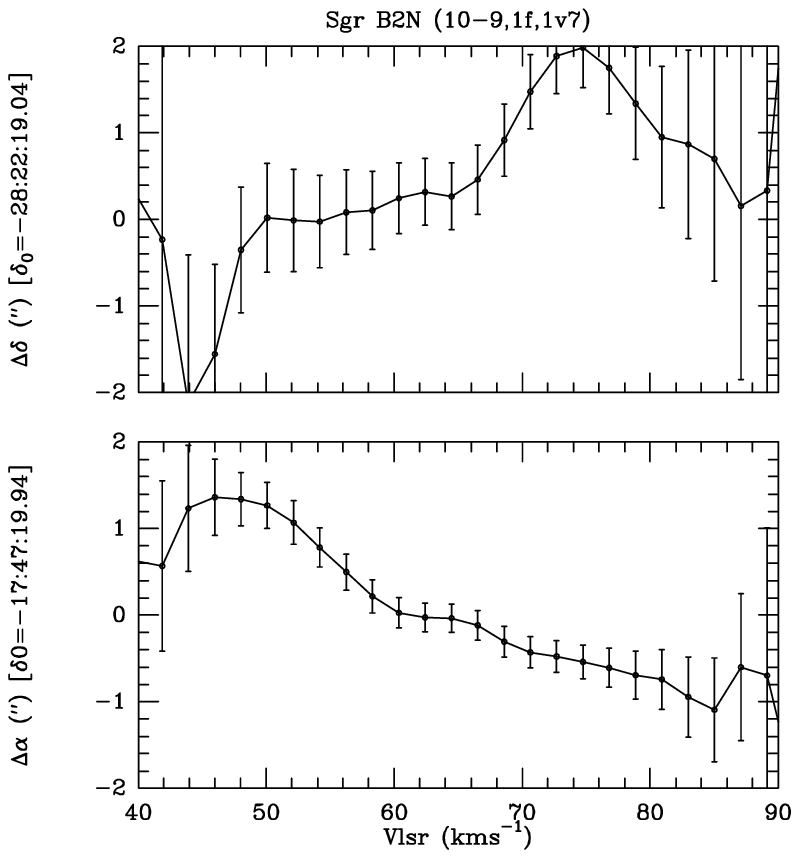}}
 \caption{10--9,1f,1v7 spatial position in RA and DEC for each velocity channel
 as obtained from a fitting process (see Sect 4). }
 \label{f:radec}
\end{figure}

 We can use the interferometric data to obtain more information about the
velocity structure in the hot cores, when the interferometric data 
have high signal to noise ratio. We have determined the relative position of 
the line emission at different radial velocities within the synthesized 
beam by fitting the relative visibilities between the individual 
velocity channels. 
Fig. \ref{f:spots} shows the location of the HC$_3$N*  
J=10--9,1v7,1f emission at different radial velocities
towards Sgr B2N and Sgr B2M superposed on
the 1.3 cm continuum image from Gaume et al. \cite*{gaume95}. We have also
included in the same plot the position of H$_2$O masers 
\cite{reid88,kobayashi89}, the NH$_3$ \cite{vogel87} and CH$_3$OH cores 
\cite{mehringer97} and our 3.3 mm peak continuum emission.
Fig. \ref{f:radec} shows an average position-velocity plot for Sgr B2N.
The data were obtained averaging the velocity channels for transitions
HC$_3$N* 10--9,1v7,1f, 1v7,1e6, 1v6,1e and 1v5,1f weighted with their error.

\subsection{Sgr B2N}

The hot gas in Sgr B2N shows a systematic trend in the position of 
the HC$_3$N* emission as the radial velocity changes. For velocity ranges 
between 
50 and 70 kms$^{-1}$ the emission at lower velocities
occurs towards the east and for higher velocities towards the west. 
For radial velocities between 70-80 kms$^{-1}$, the HC$_3$N emission 
shifts abruptly $2''$ to the north (see also Fig. \ref{f:map_pdb}, Fig. 
\ref{f:radec})
This shift suggests that the emission between 70 and
80 kms$^{-1}$ comes from a different hot core  
than the HC$_3$N emission at 60-70 kms$^{-1}$. A similar suggestion 
has also been made by Miao et al. \cite*{miao95} and Liu \& Snyder 
\cite*{liu99} from C$_2$H$_5$CN
line observations. However these authors estimate a distance of $5''$
between both peaks.

Other evidence for the presence of recent massive star formation near to 
the hot core at 70-80 kms$^{-1}$ comes from the detection of UC HII region
K3 with a radial velocity of 71 kms$^{-1}$ \cite{depree95}. Furthermore 
Lis et al. \cite*{lis93} detected extended dust emission in the north-south 
direction, which indicates the presence of two dust condensations. Therefore 
the final picture for Sgr B2N, supported by all data, 
suggests that it is composed by two hot cores, one  at 60 kms$^{-1}$ 
(Sgr B2N1) and the other at 75 kms$^{-1}$ (Sgr B2N2) separated $2''$ in a 
south-north direction.

\subsection{Sgr B2M}

HC$_3$N emission towards Sgr B2M also shows some kinematic structure 
although the larger errors make this determination very uncertain. 
Radial velocities increase for decreasing RA.  The same trend is 
observed in the H$_2$O maser emission \cite{kobayashi89}. Other molecular 
tracers show different behavior as a function of radial velocity. Vogel 
et al. \cite*{vogel87} the NH$_3$ emission found from the NH$_3$ emission 
that the blue-shifted and red-shifted components 
are separated. Gaume \& Claussen \cite*{gaume90}, with 
higher angular resolution observations of NH$_3$, see a continuous 
north-south velocity gradient. Lis et al. \cite*{lis93} point out that 
the velocity gradient obtained from HC$_3$N 25--24 emission in 
the SE-NW direction does not agree with 
the structure seen in NH$_3$ and suggest that HC$_3$N and ammonia may 
be tracing different components.


\section{Location of the emission regions}

\begin{table*}
 \caption{Location of the emission regions for the mm continuum, some 
HC$_3$N* 10--9 transitions, and other tracers towards Sgr B2M 
and Sgr B2N.}
 \begin{flushleft}
 \begin{tabular}{lrrrr}
 \multicolumn{1}{l}{} & \multicolumn{2}{c}{Sgr B2M} & 
 \multicolumn{2}{c}{Sgr B2N} \\ \hline
Transition &$\alpha$ (J2000)&$\delta$ (J2000)&$\alpha$ (J2000)&$\delta$ 
(J2000)\\
           &17$^h$ 47$^m$ &-28$^o$ 23$^m$ &17$^h$ 47$^m$ & -28$^o$ 22$^m$ 
           \\ \hline
HC$_3$N 1v7,1f      &20.15 $\pm$ 0.04&03.20 $\pm$ 0.5&19.94 $\pm$ 0.01&
  18.64 $\pm$ 0.4\\
HC$_3$N 1v7,1e      &20.15 $\pm$ 0.04&04.20 $\pm$ 0.5&19.94 $\pm 0.5$ & 
18.86 $\pm 1$\\
HC$_3$N 1v6,1e      &20.15 $\pm$ 0.07&05.20 $\pm$ 1.0&19.94 $\pm 0.5$ & 
18.86 $\pm 1$\\
HC$_3$N 1v5,1e      &20.15 $\pm$ 0.11&05.50 $\pm$ 2.0&19.94 $\pm 0.5$ & 
18.76 $\pm 1$\\ 
3.3mm Cont.         &20.25&03.60  &19.95&16.43  \\ 
3.4mm Cont. $^{(1)}$&20.17 $\pm$ 0.02&04.77 $\pm 0.5$&19.87 $\pm$ 0.02 &18.79 $\pm 0.5$ \\ 
1.3mm Cont. $^{(2)}$&20.16&04.97  &19.89&17.79 \\ 
H$_2$O  $^{(3,4)}$  &20.13 $\pm$ 0.003 &04.30 $\pm$ 0.06 &19.91 $\pm$ 0.003 &19.47 $\pm$ 0.06 \\ 
NH$_3$  $^{(5)}$    &20.17&04.77&19.88&19.09 \\
CH$_3$OH $^{(6,7)}$   &20.15 $\pm$ 0.05 &04.97 $\pm$ 0.7 &19.84 $\pm$ 0.05 &18.90  $\pm$ 0.7 \\ 
UC HII region $^{(8)}$&20.17 $\pm$ 0.01 &03.57 $\pm$ 0.1 &19.88 $\pm$ 0.01 &18.39 $\pm$ 0.1 \\ 

\hline
\multicolumn{5}{l}{\small $^{(1)}$ Carlstrom \& Vogel (1989),
$^{(2)}$ Lis et al. (1993), $^{(3)}$ Kobayashi et al (1989)} \\
\multicolumn{5}{l}{\small $^{(4)}$ Reid et al. (1989), $^{(5)}$ VGP (1987),
$^{(6)}$ Mehringer \& Menten (1997)}\\
\multicolumn{5}{l}{\small  $^{(7)}$ Houghton \& Whiteoak (1995),
$^{(8)}$ Gaume et al. (1995): F2(B2M), K2 \& K3(B2N)}
\end{tabular}
\end{flushleft} 
\label{t:loc_interf}
\end{table*}

 Table \ref{t:loc_interf} summarizes the location of the 3.4 mm continuum
and the HC$_3$N* line peak emission as derived from Gaussian fits to 
the integrated intensity of the different HC$_3$N$^*$ lines.
For comparison we have also included the positions of the H$_2$O masers 
\cite{reid88}, the 
NH$_3$ \cite{vogel87} and CH$_3$OH \cite{mehringer97} 
thermal cores, our 3.3 mm, the 3.4 mm \cite{carlstrom89} 
and 1.3 mm continuum peak emission \cite{lis93}, and the nearest 
ultra-compact HII (UC HII)
regions, K2, K3 and F2 measured at 1.3 cm \cite{gaume95}.
The position in Table \ref{t:loc_interf} for the H$_2$O maser 
emission \cite{reid88} corresponds to the cluster with 
the largest number of features (12).

The position for other masers such as H$_2$CO \cite{mehringer95}, 
OH \cite{gaume90} and 
CH$_3$OH \cite{mehringer97} detected towards Sgr B2N 
and Sgr B2M have not been included in Table \ref{t:loc_interf} because they
do not come from the same region
where the HC$_3$N* lines arise and apparently do not have relation 
with these cores. They are probably related to the hot expanding shells 
discovered by Mart\'\i n-Pintado et al. \cite*{martin-pintado99}.

As already stated in Sect 3.2, the positions of the 3.3 mm 
continuum peak are different from the HC$_3$N* peak emissions, 
indicating that
the hot cores observed in HC$_3$N are located in different regions
than the HII 
regions. Our continuum maxima at 3.3 mm also differ from the 
3.4 mm maxima \cite{carlstrom89} 
by $\sim$ 1.5$''$ and 1$''$ in Sgr B2N and Sgr B2M, which is probably
due to the complex distribution of HII regions in both sources and 
the different beam sizes. The HC$_3$N positions in Sgr B2M and Sgr B2N  
agree within $1''$ with the positions of the H$_2$O masers 
in each source, indicating that the two emissions must 
be related. Furthermore the HC$_3$N*
emission also peaks close to the CH$_3$OH and NH$_3$ thermal cores and 
therefore the two emissions are likely associated.

The IR radiation, which is the main 
excitation mechanism for HC$_3$N*, might also be responsible for the pumping
of the H$_2$O masers. The lack of radiocontinuum emission 
associated with the hot cores suggests that these IR sources have not yet 
ionized their surroundings. 

None of the Sgr B2R hot cores is coincident with any HII region
\cite{gaume95}. Sgr B2R1 is $10''$ southeast of UC HII region 
X8.33, Sgr B2R2 is $3''$ north
of UC HII region Z10.24, while Sgr B2R3 is $5''$ south of it. Sgr B2R4 is
within $1''$ distance of HII region A1. The Sgr B2W extended emission 
encompasses HII regions A2, A1, B9.89, B9.99, B9.96, Y, E, D and C 
\cite{gaume95}.
None of the Sgr B2R cores coincides with the 6.7 GHz methanol masers 
\cite{houghton95} although they are within a distance of $10''$ 
of some of them. Sgr B2R2 and Sgr B2R4 are $2''$ from two H$_2$O
maser features \cite{kobayashi89}. As for the Sgr B2M and Sgr B2N
hot cores, the Sgr B2R hot cores have not yet evolved to the UC HII 
region stage.


\section{Properties of the hot cores determined from HC$_3$N*}

 The HC$_3$N* lines observed in this paper sample hot gas in a wide range 
of temperatures from the ground state up to 1000 K. Combining 
the emission from all lines one can 
study the density and the temperature gradients in the hot cores.
 Table \ref{t:hot_cores} summarizes the main physical parameters
(excitation, temperature, HC$_3$N column density, sizes and abundance)
for the hot cores as derived from the HC$_3$N* data presented in this paper.

\begin{table*}
 \caption{Hot core properties.}
 \begin{flushleft}
 \begin{tabular}{lrrrrrrrrrrr}
 \multicolumn{1}{l}{Hot Core} & \multicolumn{1}{c}{$\alpha$ (J2000)} & 
 \multicolumn{1}{c}{$\delta$ (J2000)} & \multicolumn{1}{c}{N(HC$_3$N) $^{\rm b}$} &
 \multicolumn{1}{c}{${\rm T_{ex}}$} & \multicolumn{2}{c}{Size ($''$)} & 
 \multicolumn{1}{c}{n(H$_2$)$^{\rm b}$} & 
 \multicolumn{1}{c}{N(H$_2$)$^{\rm a}$}& 
 \multicolumn{1}{c}{X(HC$_3$N)$^{\rm b}$}& 
 \multicolumn{1}{c}{Mass$^{\rm b}$} & \multicolumn{1}{c}{L$^{\rm b}$} \\ 
        & 17$^h$47$^m$ &-28$^h$ 23$^m$& (cm$^{-2}$)  & (K) & (Lower) 
        & (Upper)    & (cm$^{-3}$)  & (cm$^{-2}$) & & (M$_\odot$)  & (L$_\odot$) \\ \hline
Sgr B2M & 20.15 & 04.00 &$7\, 10^{16}$ & 190 & 1.8$\times $1.8$^d$ 
        & 3$\times$ 5 & $5\, 10^7$ & $1.4\, 10^{25}$ &$5\, 10^{-9}$ 
        &500 & $3\, 10^6$ \\
Sgr B2N1& 19.94 & 18.74 &$10^{18}$ & 350 & 0.5$\times$0.5 
        & 3$\times$ 5 & $6\, 10^7$ & $10^{25}$ &$10^{-7}$ 
        &200$^c$ & $2\, 10^7$ \\
Sgr B2N2& 19.90 & 17.30 &$9\, 10^{16}$ & 180 & 0.5$\times$0.5 
        & 3$\times$ 5 & $2\, 10^8$ & $3\, 10^{25}$ &$3\, 10^{-9}$ 
        &600$^c$ & $3.4\, 10^6$ \\
Sgr B2R1 &18.64 & 18.74 &$5\, 10^{17}$ & 270 & 0.7 $\times$ 0.7 
         & 7 $\times$ 7 & $<10^7$ & $<2.5\, 10^{24}$ &$>2\, 10^{-7}$ 
         & $<10$& $3\, 10^6$  \\
Sgr B2R2 &20.00 & 38.00 &$7\, 10^{17}$ & 320 & 0.6 $\times$ 0.6 
         & 7 $\times$ 7 & $3\, 10^7$ & $3\, 10^{24}$ &$2\, 10^{-7}$ 
         & 30& $5\, 10^6$ \\
Sgr B2R3 &20.00 & 50.00 &$2\, 10^{17}$ & 240 & 1.0 $\times$ 1.0 
         & 7 $\times$ 7 & $<10^7$ & $<10^{24}$ &$>2\, 10^{-7}$ 
         & $<10$& $2\, 10^6$ \\
Sgr B2R4 &19.38 & 00.40 &$2\, 10^{17}$ & 250 & 0.8 $\times$ 0.8 
         & 7 $\times$ 7 & $<10^7$ & $<10^{24}$ &$>2\, 10^{-7}$ 
         & $<10$& $3\, 10^6$ \\
\hline
\multicolumn{11}{l}{\small $^{\rm a}$ Estimated
from interferometric 1.3mm flux measurements by Lis et al. (1993).} \\
\multicolumn{11}{l}{\small $^{\rm b}$ Calculated assuming $\theta_s=2''$ 
for SgrB2M and $\theta_s=1.5''$ for Sgr B2N1 and B2N2 and $\theta_s=1.5''$ 
for Sgr B2R1 to B2R4.} \\
\multicolumn{11}{l}{\small $^{\rm c}$ The total mass for the Sgr B2N core
is $\sim$ 800 M$_\odot$}\\
\multicolumn{11}{l}{\small $^{\rm d}$ Size estimated only for the 
1v7,1f and 1v7,1e emission.}\\
\end{tabular}
\end{flushleft} 
\label{t:hot_cores}
\end{table*}

\subsection{Excitation temperatures and HC$_3$N column densities}

\begin{figure}
\resizebox{\hsize}{!}{\includegraphics{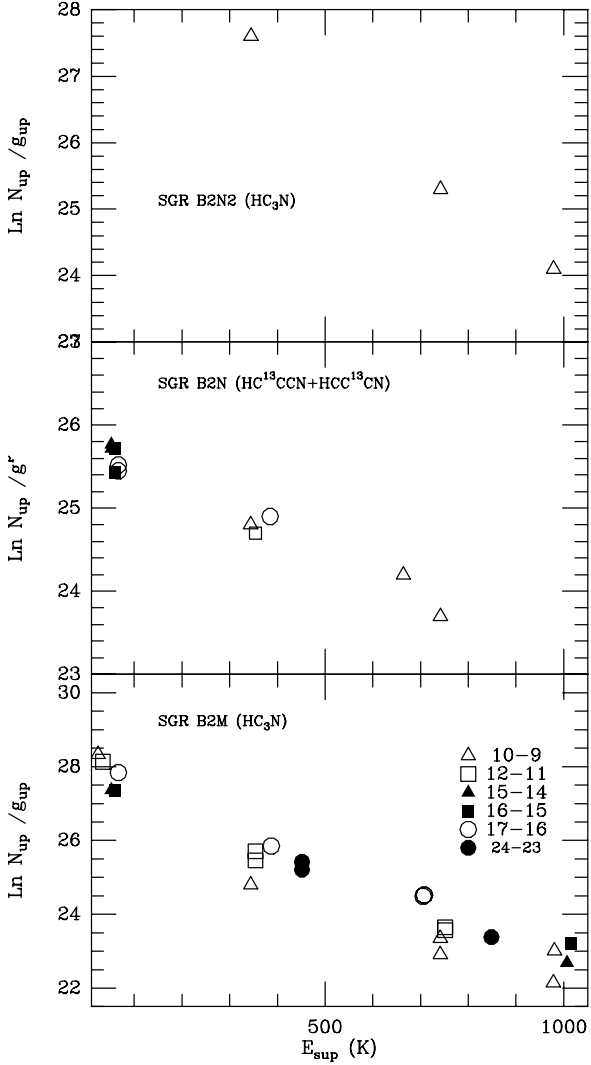}}
 \caption{Population diagram of vibrationally excited transitions.
          From single dish observations towards Sgr B2M and Sgr B2N1 and 
          from interferometric observations towards Sgr B2N2.}
 \label{f:rot_diag}
\end{figure}
 
 From the single dish optically thin HC$_3$N* lines towards Sgr B2M and the 
$^{13}$C isotopic substitution of HC$_3$N* lines towards 
Sgr B2N1 we have determined the HC$_3$N column densities in 
different vibrational levels for both hot cores. We have also used the 
interferometric HC$_3$N* lines at 75 kms$^{-1}$ to estimate the HC$_3$N column
densities in the different vibrational levels towards Sgr B2N2. The column
densities for Sgr B2N2 are uncertain because of the blending with the 
Sgr B2N1 lines and should be considered as rough guidelines.

Fig. \ref{f:rot_diag} shows the population diagram (see Goldsmith
\& Langer \cite*{goldsmith99} for an explanation of the nomenclature) 
for Sgr B2M, Sgr B2N1 and Sgr B2N2. 
Surprisingly, a straight line can fit all the data points for all sources.
The slope of this line should be related to the excitation temperature.  
We estimate excitation temperatures of $\sim$190 K towards Sgr B2M, 
$\sim$350 K from the HC$^{13}$CCN and HCC$^{13}$CN data towards Sgr B2N1 and 
$\sim$180 K towards Sgr B2N2. The excitation temperatures that we obtain 
towards 
Sgr B2M and Sgr B2N1 are in good agreement with those determined from 
the ground states lines of CH$_3$CN by de Vicente et al. \cite*{devicente97} 
and from NH$_3$ towards Sgr B2M by Vogel et al. \cite*{vogel87}.

 However, as mentioned in Sect. 4, the linewidths for the different 
vibrational transitions are different with the width of the lines decreasing 
as the energy of the vibrational levels increases. Therefore, the 
excitation temperature derived from the population diagrams does not reflect 
the actual temperature structure of the hot cores.
If we use the  
linewidth of the 1v5,1f for Sgr B2M and Sgr B2N1, and the linewidth 
of 1v6,1e for Sgr B2N2, we obtain excitation temperatures of 
$\sim$350 K for Sgr B2M, $\sim$700 K for Sgr B2N1 and $\sim$270 K for 
Sgr B2N2. This behaviour clearly indicates
that the excitation temperature in the hot core has a gradient 
as a function of the
distance to the heating source. This will be discussed in detail in the 
model presented in Sect. 7.

The beam averaged HC$_3$N column density obtained 
using excitation temperatures of 190 K towards Sgr B2M, 
350 K towards Sgr B2N1 and 180 K towards Sgr B2N2,
is $2 \, 10^{15}$ ${\rm cm^{-2}}$ for Sgr B2M,  
$10^{16}$ ${\rm cm^{-2}}$ towards Sgr B2N1 and $7\, 10^{15}$ ${\rm cm^{-2}}$ 
towards Sgr B2N2.
To obtain the total column densities we have computed the partition 
function including all vibrational levels with energies up to 1100 K, 
and assuming LTE population with the excitation temperature derived
from the rotational diagrams and a $^{12}$C/$^{13}$C ratio of 20. 

The excitation temperatures of the hot cores Sgr B2R1, B2R2, B2R3 and B2R4 
have been estimated from the J=10--9 1v5,1f, 1v6,1e and 1v7,1f 
column densities measured with the 30m dish.
In case that opacity effects are important in these lines the estimated 
excitation temperatures and column densities are lower limits to the 
actual values.
The estimated excitation temperatures are 270, 320, 240 and 250 K and the 
beam averaged HC$_3$N total column densities $10^{15}$, 
$7\, 10^{14}$, $5\, 10^{14}$ and $6\, 10^{14}$ ${\rm cm^{-2}}$
for Sgr B2R1, B2R2, B2R3 and B2R4 respectively.

\subsection{Size of the emitting regions}

The interferometric maps show that the HC$_3$N* emission towards Sgr B2N and
Sgr B2M is unresolved, and therefore we can place an upper limit 
of $3''\times 5''$ to the size of the hot cores, which is consistent 
with the size of the emission presented in Fig. \ref{f:spots}. We 
can also use 
brightness temperature arguments to estimate the area of the emitting region
when its corresponding size ($\theta_s$) is smaller than the beam 
($\theta_b$). For this,
we will use the line intensity at the radial velocity where the line peaks.
A lower limit to the area of the source, assuming circular 
geometry, is given 
by, $ \theta_s^2 \geq \theta_b^2 T_{mb} /T_b$,
where $T_b$ is the brightness temperature of the emission and $T_{mb}$ 
is the main beam temperature.
In the case of Sgr B2N1 the HC$_3$N* lines of the main isotope
are optically thick and therefore the brightness temperature must equal
the excitation temperature. Using the excitation temperature 
derived from the population diagrams ($\sim$ 350 K) and a line 
opacity of 4.5 (derived in Sect. 3.1)for 
all transitions, we obtain the source sizes summarized in Table 
\ref{t:sizes}. Columns 6 and 7 of Table \ref{t:hot_cores} give the 
upper and lower limits to the actual source sizes.
Size limits for Sgr B2N1 are consistent with the extent of this hot core 
as derived from kinematic data. 
If the lines are optically thick, the actual area of the hot cores in the
sky for the different HC$_3$N* lines must be close to those in Table 8.

\begin{table}
 \caption{HC$_3$N* core sizes towards Sgr B2N estimated from a 
 line opacity of 4.5}
 \begin{flushleft}
 \begin{tabular}{lr}
 \multicolumn{1}{l}{Transition} & \multicolumn{1}{c}{Size (arcsec)} \\ \hline
  10--9,1v7,1f & $1.5''\times 1.5''$ \\
  10--9,1v6,1e & $1.0''\times 1.0''$ \\
  10--9,1v5,1f & $0.8''\times 0.8''$ \\
  10--9,2v7,1f & $0.5''\times 0.5''$ \\
  10--9,3v7,1f & $0.7''\times 0.7''$ \\ \hline
 \end{tabular}
 \end{flushleft}
 \label{t:sizes}
\end{table}

The source size is different for the various HC$_3$N* transitions 
towards Sgr B2N1, indicating the presence of a temperature gradient. In 
this case the 
derived excitation temperature from the population diagrams 
($\sim$ 350 K) must 
be considered a mean dust temperature in the hot core. 

 The size of Sgr B2N2 is difficult to estimate since we ignore the opacity
of these lines. However from Fig. \ref{f:spots} we may accept $1.5''$ as a 
reasonable value.

 The opacities of the HC$_3$N* lines towards Sgr B2M are small 
and therefore we can only give a lower limit to the size of the core. 
The value shown in Table \ref{t:hot_cores}, $1.8'' \times 1.8''$,
has been obtained from the J=17--16,1v7,1f transition assuming that 
the line is optically thick and an excitation temperature of 190 K.
The upper limit size in Table \ref{t:hot_cores} is 
derived from our beam (1/2 HPBW).
Sizes of the Sgr B2M and Sgr B2N hot cores have also been measured 
from interferometric data, such as continuum \cite{lis93},
NH$_3$ emission \cite{vogel87} and thermal CH$_3$OH emission 
\cite{mehringer97}. These sizes are slightly larger or similar to 
those derived in this section. 
Sizes of the Sgr B2R1 to B2R4 cores are difficult to estimate. We can set 
a lower limit using the same arguments as for Sgr B2N for the 
J=24--23,3v7,1f transition. The results are summarized in Table 
\ref{t:hot_cores}.

\subsection{HC$_3$N abundance, H$_2$ density and masses}

The HC$_3$N abundance in Sgr B2M, Sgr B2N1 and Sgr B2N2 
can be inferred from 
our HC$_3$N column densities and 
the H$_2$ column density estimated by Lis et al. \cite*{lis93} 
from continuum measurements at 1.3 mm obtained with similar 
angular resolution. Since the HC$_3$N source sizes for 
the Sgr B2M, Sgr B2N1 and Sgr B2N2 hot cores are 
1-2$''$, one order of magnitude smaller than the smallest single dish 
beam used in our observations, the beam averaged HC$_3$N column density 
derived in Sect. 4.1 must 
be corrected for beam dilution effects. This increases the HC$_3$N column
density by approximately two orders of magnitude for the single dish data
and one order of magnitude for the interferometric data. The 
corrected HC$_3$N column densities for all hot cores are given in column 4 of 
Table \ref{t:hot_cores}. 

%

We have recalculated the H$_2$ column densities from the continuum measurements 
of Lis et al. \cite*{lis93}. We have assumed that the 
dust temperatures are similar to the HC$_3$N* excitation temperatures,
a dust emissivity with a power law with a slope of 1.1 for all sources 
(see Mart\'\i n-Pintado et al. \cite*{martin-pintado90} and 
Lis et al. \cite*{lis93} for a discussion on this issue), and the peak flux
densities measured by Lis et al. \cite*{lis93}, with the free-free
contribution substracted.

 Lis et al. \cite*{lis93} report a deconvolved size for the dust 
 emission towards
Sgr B2N of $<1.3''\times 5''$ in the south-north direction. 
The size of the hot core determined from the HC$_3$N* emission, 1.5$''$, 
is similar to the size of the dust in RA. As discussed in Sect. 4 
the elongation in declination of 
the dust emission is due to two hot cores with sizes of $\sim 1.5''$
separated by $2''$. We have considered half the total continuum 
flux at 1.3mm to be associated with each hot core (9 Jy/beam).

Under the previous hypothesis we obtain mean H$_2$ column densities 
of $1.4\, 10^{25}\ {\rm cm^{-2}}$ towards Sgr B2M and 
$10^{25} \ {\rm cm^{-2}}$ towards Sgr B2N1 and 
$3\, 10^{25} \ {\rm cm^{-2}}$ towards Sgr B2N2
(see Table \ref{t:hot_cores}). 
The H$_2$ column density for Sgr B2R2 has been estimated from the 
continuum interferometric flux measured by Lis et al. \cite*{lis93}, assuming  
a source size of 1 arcsec, a temperature of 350 K and a flux of 1 Jy/beam. 
Towards
Sgr B2R2 we get an H$_2$ column density of $3\, 10^{24} \ {\rm cm^{-2}}$.

For the total  
HC$_3$N column densities of $7 \, 10^{16}$ ${\rm cm^{-2}}$ for Sgr B2M, 
$10^{18}$ ${\rm cm^{-2}}$ for Sgr B2N1,  
$9\, 10^{16}$ ${\rm cm^{-2}}$ for Sgr B2N2
and $7\, 10^{17}$ for Sgr B2R2
and the H$_2$ calculated densities, we determine a fractional 
HC$_3$N abundance of $5\, 10^{-9}$, $10^{-7}$, $3\, 10^{-9}$ and $2\, 10^{-7}$ 
towards Sgr B2M, Sgr B2N1, Sgr B2N2 and Sgr B2R2 respectively.

 For the other hot cores in the ridge we can only set lower limits to 
the HC$_3$N
abundance considering the upper limit to the continuum flux density at 1.3mm
and assuming a typical size for the hot cores of $1''$. We estimate a
lower limit to the HC$_3$N abundance of $2\, 10^{-7}$, in agreement
with that obtained for the Sgr B2R2
hot core. Lower abundances of HC$_3$N are possible only if the hot cores 
Sgr B2R1, B2R3 and B2R4 are substantially larger.

For a depth along the line of sight similar to the size of the hot cores
we estimate mean H$_2$ densities of $7\, 10^7\ {\rm cm^{-3}}$ for 
Sgr B2M and $6\, 10^7\ {\rm cm^{-3}}$ for Sgr B2N1, 
$2\, 10^8\ {\rm cm^{-3}}$
for Sgr B2N2 and 
$3\, 10^7 {\rm cm^{-3}}$ for Sgr B2R2. Assuming a spherical geometry, 
and the previous densities we derive masses of 
500 M$_\odot$ for Sgr B2M and 200 M$_\odot$ for Sgr B2N1, 
and 600 M$_\odot$ for Sgr B2N2. Then the 
total mass in Sgr B2N would be roughly $\sim$ 800 M$_\odot$, similar 
to that in Sgr B2M. The masses for the
ridge hot cores are between 10 to 30 M$_\odot$, substantially
smaller than those for Sgr B2M and Sgr B2N (see Table \ref{t:hot_cores}).

\subsection{Luminosities}
 The high temperatures obtained from the HC$_3$N* emission indicate that the
hot cores are internally heated.
For the derived H$_2$ densities ($10^7 - 10^8\ {\rm cm^{-3}}$) 
gas and dust are coupled and the excitation temperature given in Table
\ref{t:hot_cores} should also correspond to the dust temperature.

For the H$_2$ column densities derived previously, the dust is optically 
thick even at 100 $\mu$m, indicating that the hot cores emit 
like a blackbody at the dust temperature. The bolometric luminosity of 
the hot cores can be estimated from the Stefan-Boltzmann law.
The results, obtained assuming a source radius of $1''$ for Sgr B2M,  
$0.75''$ for Sgr B2N1 and Sgr B2N2 and $0.5''$ for Sgr B2R1 to B2R4, 
are summarized in Table \ref{t:hot_cores}. 

%

The luminosity may be overestimated due to geometric effects (the
sources may have a disk geometry instead of a spherical one, see 
Cesaroni et al. \cite*{cesaroni97}). However lower luminosities would 
not explain the sizes of 
the HC$_3$N* emission lines (see next section). 

\section{The distribution of the hot dust in the hot cores}

Goldsmith et al. \cite*{goldsmith85} argued, based on the 
thermalization at only one temperature and on the large H$_2$ 
densities ($10^{10}\ {\rm cm^{-3}}$) required for the 
collisional excitation of the HC$_3$N* lines, that these lines
are excited by IR radiation. Similar arguments also apply to the Sgr B2 hot 
cores. The HC$_3$N* rotational lines from different vibrational levels 
will trace IR emission at the wavelength of the vibrational transitions.
One can then use the different sizes obtained from the
lines to study the thermal 
structure of the dust and gas surrounding the young massive stars.

In order to understand the population diagrams derived from the HC$_3$N* data
we have developed a simple model to study the excitation 
of the HC$_3$N* emission for a condensation of gas and dust surrounding 
a young star. For simplicity, we have considered that the hot cores have 
spherical geometry with an inner radius
of $10^{16}$ cm. The excitation temperature of the HC$_3$N* lines 
(1v7, 1v6, 1v5, 2v7 and 3v7) has been assumed to be 
equal to the dust temperature as expected from H$_2$ 
densities of $>10^6$ cm$^{-3}$ derived for the hot cores in Sgr B2. 
For the H$_2$ column densities we derive for the hot cores 
(Table \ref{t:hot_cores}), 
the dust emission at the wavelenghts where the
vibrational excitation of HC$_3$N ocurrs is optically thick. In this case, one
can consider that dust and gas are coupled and that
the radial profile of the dust temperature can be approximated
by the Stefan-Boltzmann law:

\begin{eqnarray}
T_{dust} = \left(\frac{L_\star}{4\pi\sigma r^2}\right)^{1/4} =
15.222 \left( \frac{L_\star}{L_\odot} \left[\frac{10^{16} 
{\rm cm}}{r}\right]^2 \right)^{1/4}
\end{eqnarray}

%
%

This approximation gives a similar dust temperature profile
to that obtained by Lis \& Goldsmith \cite*{lis90} 
from models that solve the radiative transfer in a dusty cloud
for the distances from the heating source relevant to our model, 
between $3\, 10^{16}$ cm and $3\, 10^{18}$ cm.
The main discrepancies between the temperatures from expression 1 and the 
radial profile from Lis \& Goldsmith \cite*{lis90}, are found in a very 
small region ($10^{16}$ to $3\, 10^{16}$ cm) around the
star which has a negligible contribution to the HC$_3$N* line intensities.

To perform the radiative transfer in the cloud we have divided the hot 
core into 300 shells. 
The H$_2$ density has been considered to decrease with the distance as 
$r^{-3/2}$ as expected for a hot core \cite{osorio99}. We have also considered
a constant HC$_3$N abundance through the hot core. 
The widths of the HC$_3$N* 
lines used in the model were 6 and 10 kms$^{-1}$ for Sgr B2M and Sgr B2N1 
respectively. 

\begin{figure}
\resizebox{\hsize}{!}{\includegraphics{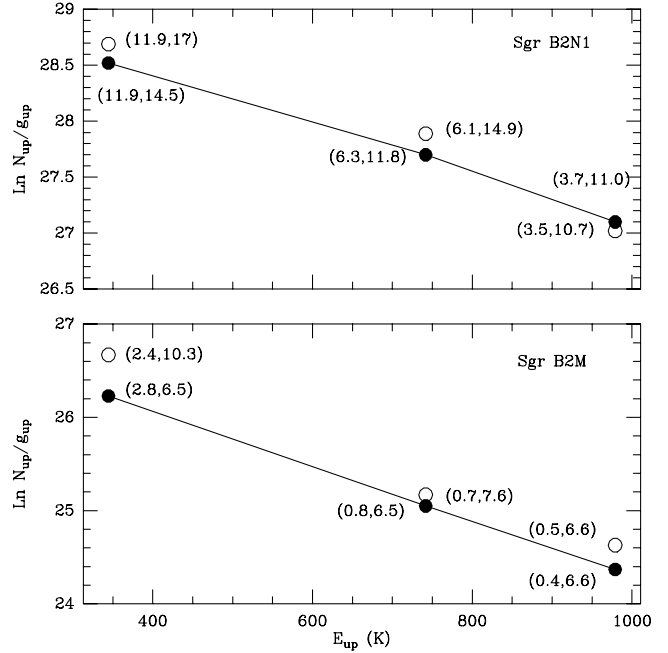}}
 \caption{Population diagrams comparing the interferometric data (open 
 circles) and the results from the model for transitions 10--9 1v7,1f, 
 1v6,1e and 1v5,1f (filled circles). Numbers between parenthesis indicate i
 T$_{mb}$ (K) and line width (kms$^{-1}$ respectively for each transition.}
\label{f:modelresults}
\end{figure}

Table \ref{t:model} shows the parameters for Sgr B2M and Sgr B2N1 
that give the best fit to the HC$_3$N* data. 
Fig. \ref{f:modelresults} shows the model predictions in the form of 
population diagrams. To compare the results of the model (filled circles) 
with our data we also show in Fig. \ref{f:modelresults} a comparison 
with the population diagrams derived from the data obtained 
with the interferometer (open circles). 
It is interesting to note that, in spite of the presence of a temperature 
gradient within the hot core, as expected for excitation by radiation, 
the population diagrams shows a straight line indicating a single 
excitation temperature. 

\begin{figure}
\rotatebox{-90}{\includegraphics{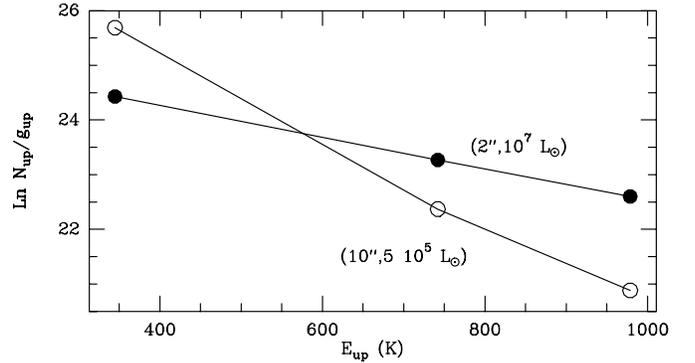}}
 \caption{Population diagram obtained from the model described 
 in Sect. 7, comparing a source of a size of 
 $2''$ with a luminosity of $10^7$ L$_\odot$, (filled circles), with 
 source of a size of $10''$, with a luminosity of $5\, 10^5$ L$_\odot$ (open
 circles).  The points correspond to transitions 10--9, 1v7,1f, 
 1v6,1e and 1v5,1f.}
\label{f:linearecta}
\end{figure}

Although the sizes of the hot cores have been estimated from the HC$_3$N* 
line intensities we have considered both the sizes and the luminosities as
free parameters. One interesting result from our modelling is that the 
shape of the population diagrams depends on the size and the luminosity 
of the hot core. Fig. \ref{f:linearecta} shows the population diagrams for 
a hot core with a size of $2''$ as measured for Sgr B2N1 with the estimated 
luminosity of $10^7$ L$_\odot$ and a hot core with the size of $10''$ 
and a luminosity of
$5\, 10^5$ L$_\odot$. The population diagram predicted for the large hot core
with low luminosity shows the typical shape of a temperature gradient and cannot
be fitted by a straight line. On the other hand, one obtains the population 
diagram that can be fitted by a straight line if the size of the hot core 
is relatively small and the luminosity high. This is because for the latter 
case the range of dust temperatures within the hot cores is relatively
small; the range of temperatures that contribute to the HC$_3$N* emission 
is between 200 and 330 K (radii between $1.8\, 10^{17}$ and $7\, 10^{16}$ cm). 
Under this condition
the population diagrams will be represented by a straight line with an 
excitation temperature close to the average temperature in the hot core.
This indicates that the HC$_3$N* emission should be restricted to a smaller 
region to that expected to be excited in the 1v7 line. The small size
of the HC$_3$N* hot core could be due to a sharp drop of the
HC$_3$N abundance or of the H$_2$ densities. The first possibility is the most
likely since the large HC$_3$N abundances (see Sect. 8) requires a particular
chemistry (HC$_3$N evaporation from grains) which is strongly coupled to 
the temperature.

\begin{table}
 \caption{Final parameters used in the model.} 
 \begin{flushleft}
 \begin{tabular}{lrr}
 \multicolumn{1}{l}{Parameter} & \multicolumn{1}{c}{Sgr B2M} 
 & \multicolumn{1}{c}{Sgr B2N1} \\ \hline
Luminosity (L$_\odot$)           &   $2\, 10^7$       &       $10^7$       \\
Distance (pc)                    &      7100          &            7100    \\
n(H$_2$) (cm$^{-3}$, $r_0=10^{16} $cm)
                                 &$6\, 10^8 (r/r_0)^{-3/2}$ 
                                                & $10^9 (r/r_0)^{-3/2}$    \\
HC$_3$N line width (kms$^{-1}$)  &         6.5        &              10    \\ 
HC$_3$N abundance                &$5\, 10^{-9}$       & $10^{-7}$          \\ 
Source size (arcsec)             &         3          &             1.9    \\
\hline
 \end{tabular}
 \end{flushleft}
 \label{t:model}
\end{table}

 The main results obtained from the model can be summarized as follows:

\begin{itemize}
\item The observed line intensities (values in parenthesis in Fig. 
\ref{f:modelresults}) towards Sgr B2N1 and Sgr B2M are fitted by the model
within 20\%.
The discrepancies, between the model predictions and the data in the 
population diagrams are due to the different line width of the observed
lines.

\item The model predicts that the lines towards Sgr B2N1 are broadened by 
opacity.  However the opacity cannot reproduce the observed large linewidth
for the 1v7 and 1v6 transitions. This is consistent with the HC$^{13}$CCN* 
data and indicates the presence of systematic kinematic effects within the
hot core.

\item The luminosity needed to fit the line intensities towards Sgr B2M is 
higher than that estimated in Sect. 6.4 by a factor of $\sim$10. If we use 
the luminosity in Table \ref{t:hot_cores}, the observed $T_{mb}$ ratios between 1v7, 1v6 and 1v5 
transitions cannot be reproduced.

\end{itemize}

\section{Hot core chemistry}

One of the most remarkable results from our data is the large abundance
of HC$_3$N towards the hot cores in Sgr B2, specially towards Sgr B2N1 and
Sgr B2R1 to B2R4, where it is of the order of $10^{-7}$. Typically, 
HC$_3$N emission is found in warm 
clouds like the extended ridge and fingers in Orion A \cite{rodriguez-franco98}
where its abundance is $10^{-9}$. In Orion A, the
HC$_3$N abundance increases by one order of magnitude ($10^{-8}$) for the
hot core and decreases one order of magnitude towards the photo-dissociation
regions in the ionization fronts of the HII region Orion A. In fact, 
Rodr\'\i guez-Franco et al. \cite*{rodriguez-franco98} propose that the 
large abundance of HC$_3$N 
emission is a good tracer of hot cores and regions which are well shielded 
from the UV radiation.
The HC$_3$N abundance we estimate for Sgr B2N1 and 
Sgr B2R1-B2R4 agrees with the maximum abundance in the gas phase 
that Caselli et al. \cite*{caselli93} obtain, at a time of 
$\sim 3\, 10^4$ yr, after 
the startup of gravitational collapse for an Orion hot core type. Caselli
et al \cite*{caselli93} developed a model that considers the evolution of 
the chemistry of some molecules in the formation of a massive star.
The HC$_3$N abundance towards Sgr B2M is better fit by the abundance values 
obtained for the hot core phase, when accretion has stalled and 
molecules are evaporated from grains.

The large difference in the HC$_3$N abundance between Sgr B2M and 
Sgr B2N1 
could also be related to the UV radiation from nearby 
OB stars that penetrates deeper in the hot core and photodissociate the 
fragile molecules like HC$_3$N. Indeed the Sgr B2M complex 
has more than 20 HII regions powered by OB stars within 5$''$ from
the Sgr B2M hot core \cite{gaume95} which should produce an intense UV 
radiation field. This contrasts with Sgr B2N, where there are fewer 
HII regions.
The same explanation applies to the Sgr B2R cores, where the abundance
is similar to that in Sgr B2N and where there is one or no HII regions 
in the neighborhood of the hot core. The low HC$_3$N abundance 
towards Sgr B2N2 has no simple explanation. 

\section{Kinematics of the Sgr BM and Sgr B2N hot cores}

One of the most remarkable results of the observations of HC$_3$N* towards 
the hot cores is the decrease of the linewidth as the energy of the 
vibrational level in which the rotational lines arise increases. This 
systematic trend in the linewidth combined 
with different sizes of the region where the different vibrationally excited 
lines arise indicates that the linewidth of the HC$_3$N* lines depends on the
distance to the exciting source. There are two possibilities to explain 
the larger line widths for the larger distances. 

The first possibility is that 
the HC$_3$N* linewidth is dominated by turbulence, with the 
linewidth depending on the size of the eddies.
The second possibility is that the velocity systematically increases with 
the distance to the source. This is the most likely explanation in view of the
previous kinematical results on both sources. We discuss now the implications
of the results presented in this paper on the kinematics of the hot cores in
Sgr B2M and Sgr B2N.

\subsection{Sgr B2N}

The kinematics in the Sgr B2N hot core has been subject of debate. 
Lis et al. \cite*{lis93}, from the map of the J=25--24 HC$_3$N line, 
and de Vicente \cite*{devicente94}, based on the spatial distribution of 
the high velocity gas observed in the J=13--12 and J=20--19 OCS 
rotational lines, proposed an east-west outflow.
Reid et al. \cite*{reid88} suggested from a model to fit the 
proper motion of the 
H$_2$O masers, that the kinematics is due to a solid body rotation 
in the E-W direction together with a spherical expansion. On the
other hand Vogel et al. \cite*{vogel87} and Kuan \& Snyder \cite*{kuan94} 
report 
a southeast-northwest velocity structure in NH$_3$ and HNO respectively.

The presence of two hot cores with different radial velocities clarifies 
the inconsistencies observed in different tracers. 
As already discussed in Sect. 4, the final picture for the emission
towards Sgr B2N as two hot cores, Sgr B2N1 and Sgr B2N2 at $\sim 60$ and 
$\sim 75$ kms$^{-1}$, respectively, separated by $2''$ in the 
north-south direction, explains the north-south velocity gradient 
observed in molecules like NH$_3$ and HNO.
However the east-west velocity gradient observed in HC$_3$N, OCS and the 
H$_2$O masers is unclear. From the velocities and proper motions of the 
H$_2$O masers a solid rigid disk has been proposed. This is also consistent
with the presence of a kinematical structure in the source as described in 
Sect. 7 to explain the difference in the line widths.
If the HC$_3$N* data is to be explained
by a solid rigid disk, the dynamical mass needed to bind the material 
in a disk which is inclined with respect to the line of 
sight by an angle $i$, and whose maximum observed velocity is $v_{max}$
is given by:
\begin{eqnarray}
M [M_\odot]= 232.49 \, (v_{max}[km/s])^2 \, r [pc] \frac{1}{\sin^2i} 
\end{eqnarray}
 For the inclination derived from the kinematics and proper motions of 
H$_2$O masers \cite{reid88}, and a maximum radial velocity of 
9 km/s at a distance of $0.8''$ (0.027 pc) from the center of the source,  
the dynamical mass required to bind the disk is 750 M$_\odot$. 
This is nearly a factor 3 larger than the mass we derive in Sect. 6.3.
However, the analysis of the proper motions and kinematics 
from the H$_2$O masers
is based on the fact that the emission comes from one source, while the 
HC$_3$N* data clearly shows that there are two hot cores.
Another possibility is that 
HC$_3$N* may come from a cluster of 
hot cores with typical masses of $\sim 20$ M$_\odot$, similar to that in 
Orion A, which have been formed from molecular gas with a 
velocity gradient. The star formation may have been triggered by the 
nearby HII regions and the hot cores now reflect the kinematics of these 
regions. In fact, the morphology of the hot cores in Fig. \ref{f:spots} 
resemble that of the southern edge of HII region K5. Higher angular 
resolution observations are needed to distinguish between the two different
scenarios.

\subsection{Sgr B2M}
As stated for Sgr B2N the trend of HC$_3$N* line widths in Sgr B2M 
would also be consistent with the kinematics of a solid rigid.
The dynamical mass required to bind the material in this case 
would be $\sim$ 400 M$_\odot$, 
which is similar to the mass we estimated in Sect. 6.3.
It may also be possible that the kinematics may be related to the 
velocity structure of multiple hot cores being formed in Sgr B2M.
Higher angular observations are required to settle this discussion.

\section{Global view of massive star formation in Sgr B2}

Sgr B2 has been extensively studied in the radio continuum and in line 
molecular emission with very high angular resolution and sensitivity. 
The number of HII regions detected in Sgr B2 is nearly 60 \cite{gaume95} 
whereas the number of detected hot cores is only 2. This difference in 
the number of detections has been proposed to be due to 
the different 
lifetimes of the UC HII region phase ($10^5$ yr, Wood \& Churchwell 1989)
and the hot core phase ($10^3 - 6\, 10^4$ yr, Kurtz et al. 2000).
However the detection of typical hot cores like those in Orion either 
using molecular lines
or dust emission requires better sensitivity than detection of UC HII regions
This indicates that a complete inventory of recent massive star formation 
requires very sensitive observations of molecular lines or/and dust 
to detect all the hot cores in the region.

 The results presented in this article modify the view of very
recent massive star formation in the Sgr B2 cloud. It has been considered that
massive stars were only forming in the main cores Sgr B2M and Sgr B2N and to
a less extent in another core to the south, Sgr B2S. This view is 
basically based on the radiocontinuum emission and on the maser activity in 
the region. 

 We have detected four new hot cores (Sgr B2R1-B2R4) in a ridge 
connecting Sgr B2M and B2N and of a extended HC$_3$N$^*$ emission to the 
west of Sgr B2M (Sgr B2W). The latter, when observed with higher 
angular resolution
and better sensitivity, may split into even larger number of
individual hot cores. This indicates
that massive star formation in the Sgr B2 cloud 
is ongoing in the large envelope at stages previous to the HII region 
phase. Similar results have also been found by 
Mart\'\i n-Pintado et al. \cite*{martin-pintado99}
in the southern part of the envelope of Sgr B2.  
The new hot cores reported in this paper, B2R1 to B2R4, and also 
those of Mart\'\i n-Pintado \cite*{martin-pintado99} 
are less massive than those in  Sgr B2M and B2N but they are similar to 
those found in Orion and other regions with massive star formation. 

 Taking into account all data we can distinguish three star 
burst forming regions with massive stars at different stages 
of evolution, in Sgr B2M, Sgr B2N and the Sgr B2R 
(the ridge). The differences in evolution among these regions arise from 
their luminosities and the ratio between the number of HII regions and the 
number of hot cores.

 We will refer to the Sgr B2M cluster as the region containing the F and G 
UC HII regions \cite{gaume95}. This cluster contains 23 
UC HII regions, one massive hot core ($\sim$ 500 M$_\odot$)
with a luminosity of $3\, 10^6$ L$_\odot$ 
and a large H$_2$O maser luminosity. The abundance 
of HC$_3$N towards this core is 
20 times lower than towards Sgr B2N.

 The Sgr B2N cluster encompasses 7 HII regions \cite{gaume95}, and 
contains two luminous hot cores of $2\, 10^7$ and  $3\, 10^6$ L$_\odot$
with a total mass of 800 
M$_\odot$ and several H$_2$O masers. The Sgr B2N1 hot core shows 
an enhanced abundance of HC$_3$N. From the results of Sect. 7, 
the dust around the core is optically thick.

 The Sgr B2 Ridge cluster is a scattered group of hot cores detected in the 
HC$_3$N* lines, with similar luminosities to the previously-mentioned
hot cores, but with
a much lower masses (10-30 M$_\odot$). These hot cores are close 
to the HII regions X, Z, and A. Sgr B2W is also close 
to HII regions E, D, C, B, A and Y \cite{gaume95}. Only Sgr B2R2 and 
Sgr B2R4 seem 
to be associated with H$_2$O masers. The HC$_3$N abundance is similar 
to that found in Sgr B2N.

The fact that the Sgr B2M and Sgr B2N hot core emission is associated
with two dense clusters of UC HII regions and that the H$_2$ column 
density shows its highest value in the region supports the hypothesis 
that these two massive hot cores may be forming a new cluster of 
massive stars. Arguing that a typical 
hot core with luminosities $L\simeq 5\, 10^5$ L$_\odot$, like Orion, 
has a mass of $\simeq$ 10-20 $M_\odot$,
the number of Orion-type hot cores in Sgr B2M and Sgr B2N
(massive stars) to be found must be 20-30. This would be close to the number
of HII regions already formed in the Sgr B2M cluster.
The difference in the HC$_3$N abundance between both hot cores is probably 
due to the fact that the cluster of massive stars in Sgr B2N is at 
an earlier stage than that in Sgr B2M.

The Sgr B2R hot cores and the Sgr B2 W extended emission come from a
region where HII regions are isolated or form groups of 2 or 3 individuals
and the H$_2$ column density is 5-10 times lower than in the other two
massive hot cores.
These new hot cores probably represent formation of more or
less isolated massive stars. However the hot cores must contain 
stars of similar mass and luminosity to those found in the Sgr B2M 
and Sgr B2N regions. The similar abundance of HC$_3$N in the Sgr B2 Ridge 
cores and Sgr B2N hot cores would suggest that they are at a similar stage 
of evolution.

Higher angular resolution observations with better sensitivity are 
necessary to reveal the number of massive stars being formed in this
molecular cloud and its relation to the physical conditions 
of the surrounding material.

\section{Conclusions}
We have used the IRAM 30m telescope and the IRAM Plateu de Bure 
interferometer to study
the hot cores at the Sgr B2 molecular cloud using several vibrationally excited 
HC$_3$N transitions. The main results derived from this work are the following:

\begin{enumerate}
  \item A ridge of HC$_3$N* emission has been discovered between Sgr B2M 
  and Sgr B2N. We distinguish four new hot core sources in the ridge 
  labelled as Sgr B2R1, B2R2, B2R3 and B2R4. The map also 
  shows an extended low brightness emission to the west of the ridge which
  we refer to as Sgr B2W.

  \item The emission from the main hot cores, Sgr B2M and Sgr B2N is unresolved
  in our interferometric maps
  and we set an upper limit to the size of the cores of $3''\times 5''$. The 
  upper limit to the size for the ridge hot cores is $7''\times 7''$.
  
  \item By fitting the relative visibilities between velocity channels 
  in our interferometric map we have obtained the relative 
  location of the HC$_3$N* J=10--9,1v7,1f emission for different radial 
  velocities towards Sgr B2N and Sgr B2M. We propose that Sgr B2N is composed
  of two hot cores separated $2''$ in the south-north direction, with radial
  velocities of 60 (Sgr B2N1) and 75 kms$^{-1}$ (Sgr B2N2).
  The estimated size for both hot cores is $\sim 1.5''$.

  \item The Sgr B2M hot core shows a velocity gradient. The radial velocity
  increases for decreasing RA. The estimated size for the hot core is 
  $\sim 2''$.

  \item The linewidths of the HC$_3$N* lines arising from different 
  vibrational
  excited states systematically decreases for Sgr B2M and Sgr B2N1 as
  the energy of the vibrationally excited states increases. This kinematical
  effect may be due to a rotating solid rigid disk. 

  \item We have estimated the kinetic temperature, H$_2$ density, HC$_3$N 
  abundance, mass and luminosity of the hot cores. 
  Typical luminosities are $2 - 3\, 10^6$ 
  L$_\odot$ for Sgr B2M and the B2R cores and $2\, 10^7$ L$_\odot$ for 
  Sgr B2N. The estimated masses for Sgr B2M and B2N are 500 and 800
  M$_\odot$ respectively and 
  $10-30$ M$_\odot$ for Sgr B2R1 to B2R4.
  The abundance of HC$_3$N towards Sgr B2M is $5\, 10^{-9}$ while 
  towards Sgr B2N1 and B2R1 to B2R4 is $1 - 2\, 10^{-7}$. 

  \item From the HC$_3$N* line intensities we have modeled the spatial
  distribution of the hot dust in the hot cores Sgr B2M and Sgr B2N. 
  The different HC$_3$N abundance between Sgr B2M and Sgr B2N is probably 
  related to the penetration of UV radiation into the hot cores and/or to
  time dependent chemistry. We 
  propose that HC$_3$N abundance probes the evolutionary stage of the hot core. 

  \item We find that the excitation of the HC$_3$N* emission in
  all hot cores can be represented by a single temperature.
  We have developed a simple model to study the excitation
  of the HC$_3$N vibrational levels by IR radiation.  We find that the single
  excitation temperature can be explained by high stellar luminosities
  ($\sim 10^7$ L$_\odot$ ) and small source sizes ($\sim 2-3''$).

  \item The Sgr B2M and B2N hot cores are likely forming a new cluster of 
  massive 
  stars. If the typical mass of a hot core like Orion has 20 M$_\odot$ 
  the number of Orion-type hot cores
  (massive stars) to be found in Sgr B2M and Sgr B2N must be 20, close to 
  the number of HII regions already formed in the Sgr B2M and Sgr B2N cluster.

  \item The Sgr B2R hot cores probably represent formation of isolated 
  massive stars. However the hot cores must contain stars of similar
  mass and luminosity to those found in Sgr B2M and Sgr B2N.
\end{enumerate}

\acknowledgements
We would like to greet the IRAM staff, and operators for their
help during observations. We are also very grateful to Dr. 
P.F. Goldsmith, for a critical reading which substantially improved 
this paper.
This work has been partially supported by the 
CYCIT and the PNIE under grants PB96-104, 1FD97-1442 and ESP99-1291-E.

\bibliography{aamnem99,ms9779}
\bibliographystyle{aabib99}

\end{document}